\documentclass{SciPost}

\hypersetup{
    colorlinks,
    linkcolor={red!50!black},
    citecolor={blue!50!black},
    urlcolor={blue!80!black}
}

\usepackage[bitstream-charter]{mathdesign}
\DeclareSymbolFont{usualmathcal}{OMS}{cmsy}{m}{n}
\DeclareSymbolFontAlphabet{\mathcal}{usualmathcal}

\fancypagestyle{SPstyle}{
\fancyhf{}
\lhead{\colorbox{scipostblue}{\bf \color{white} ~SciPost Physics Community Reports }}
\rhead{{\bf \color{scipostdeepblue} ~Submission }}

\fancyfoot[C]{\textbf{\thepage}}
}

\begin{document}

\pagestyle{SPstyle}

\begin{center}{\Large \textbf{\color{scipostdeepblue}{
Model-Agnostic Signal Discovery with Machine Learning:\\Bridging the Gap Between Theory and Practice
}}}
\\
\medskip
Part of the VERaiPHY initiative
\end{center}

\begin{center}\textbf{
Oz Amram\textsuperscript{1$\star$},
Marco Letizia\textsuperscript{2,3$\dagger$} and
Mikael Kuusela\textsuperscript{4$\ddag$}
}
\end{center}

\begin{center}

{\bf 1} Fermi National Accelerator Laboratory, Batavia, IL 60510, USA
\\
{\bf 2} MaLGa-DIBRIS, Università di Genova, Via Dodecaneso 35, I-16146 Genoa, Italy
\\
{\bf 3} INFN, Sezione di Genova, Via Dodecaneso 33, I-16146 Genoa, Italy
\\
{\bf 4} Department of Statistics and Data Science, Carnegie Mellon University, Pittsburgh, PA 15213, USA
\\[\baselineskip]
$\star$ \href{mailto:oz.amram@cern.ch}{\small oz.amram@cern.ch} 
$\dagger$ \href{mailto:marco.letizia@edu.unige.it}{\small marco.letizia@edu.unige.it}
$\ddag$ \href{mailto:mkuusela@andrew.cmu.edu}{\small mkuusela@andrew.cmu.edu} 
\end{center}

\section*{\color{scipostdeepblue}{Abstract}}
\textbf{\boldmath{
Searches for new phenomena in complex scientific data are predominantly model-dependent, optimized for specific hypotheses, and therefore limited in their coverage of the space of possible signals. Recently, new AI-based model-agnostic search strategies, many of which have been pioneered in high-energy physics, have been proposed which provide a complementary paradigm, prioritizing broad exploration over tailored analyses. These techniques offer an opportunity to enhance the overall discovery potential of modern experiments, especially in regimes where theoretical guidance is scarce. In this document, we review the conceptual framework behind the main classes of AI-based model-agnostic strategies. We discuss the potential pitfalls of these methods, and strategies for their validation and interpretation. We aim for this document to serve as a useful reference both for practitioners and for researchers interested in learning more about these model-agnostic search strategies.
}}

\vspace{\baselineskip}

\noindent\textcolor{white!90!black}{%
\fbox{\parbox{0.975\linewidth}{%
\textcolor{white!40!black}{\begin{tabular}{lr}%
  \begin{minipage}{0.6\textwidth}%
    {\small Copyright attribution to authors. \newline
    This work is a submission to SciPost Phys. Comm. Rep. \newline
    License information to appear upon publication. \newline
    Publication information to appear upon publication.}
  \end{minipage} & \begin{minipage}{0.4\textwidth}
    {\small Received Date \newline Accepted Date \newline Published Date}%
  \end{minipage}
\end{tabular}}
}}
}


\vspace{10pt}
\noindent\rule{\textwidth}{1pt}
\tableofcontents
\noindent\rule{\textwidth}{1pt}
\vspace{10pt}


\section{Introduction}

One of the primary research activities in the field of high-energy physics (HEP) is the analysis of large datasets to search for evidence for new phenomena.
To accommodate the large volume and high complexity of the data, and achieve gains in sensitivity, searches for new phenomena often restrict their scope to specific models or classes of new particles.
Other approaches instead attempt to make minimal assumptions and probe a broad range of possibilities, but achieve reduced sensitivity as compared to dedicated searches. 
This gives rise to two families of methods: \emph{model-dependent} and \emph{model-agnostic} (or \emph{model-independent}) approaches.

Model-dependent searches are designed on the basis of well-specified background and signal hypotheses. 
They look for deviations from a background model while incorporating properties of hypothetical signals (such as particle masses, decay modes, or kinematic distributions) and optimize the analysis to enhance sensitivity to these signatures, for example by focusing on a few relevant high-level features such as an invariant mass. 
For any specific hypothesis to be tested, these methods are generally the most powerful. 
Indeed the likelihood-ratio test, used ubiquitous in HEP data analyses, is guaranteed by the Neyman--Pearson lemma \cite{Neyman:1933wgr} to be the most powerful test for fully specified background and signal hypotheses. 
However, their sensitivity is typically limited to a narrow region of the whole space of possible new physics scenarios and observables, and they heavily rely on high-fidelity models for the background processes.

\begin{figure}
    \centering
    \includegraphics[width = 0.5 \textwidth ]{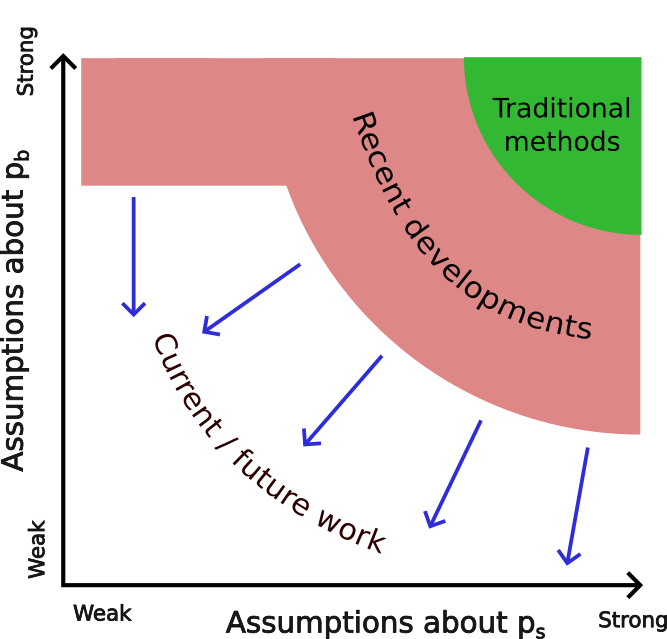}
    \caption{Landscape of model-agnostic signal detection. The various methods can be categorized according to their strength of assumptions about the background distribution ($p_b$) and the signal distribution ($p_s$).}
    \label{fig:landspace}
\end{figure}

In contrast, model-agnostic searches are designed to minimize assumptions about the signal and/or the background. 
It is in fact useful to distinguish between background- and signal-agnostic methods as illustrated in Figure~\ref{fig:landspace}. 
These approaches are valuable when there is little guidance from theory or when predictions via Monte Carlo simulations are not particularly accurate or reliable. 
In this review we will focus on the case where minimal assumptions are being made about the signal hypothesis. Different techniques will be required depending on the degree to which the background hypothesis is well specified. 
While model-agnostic methods enable exploration of a broader spectrum of potential new physics signatures, they lack the sensitivity of model-dependent strategies for specific signals and often require careful statistical validation to control false discovery rates. 
They also face crucial challenges in their interpretation, both in the case of a significant discrepancy and in the reporting of exclusion limits. Furthermore, it is worth highlighting that a model-agnostic search is still typically restricted to a subset of final states and observables that are physically motivated or experimentally accessible. 

In practice, most analyses fall somewhere between these two extremes. For example, bump hunts \cite{Choudalakis:2011qn}, which search for localized excesses, can be considered partially model-independent: they do not depend on specific particle-physics models and often rely on data-driven estimates of the background  \cite{CMS:higgs,ATLAS:higgs,CMS:dijet,ATLAS:dijet}, yet they still assume that any signal will appear as a localized excess in a particular variable (for example, an invariant-mass distribution) within a predefined signal region. Another approach is to consider a comprehensive enumeration of final states, typically hundreds to thousands, while defining a small set of observables for each, and compare the observed data in each final state to simulations of the background processes. 
One then defines a statistical procedure to quantify the largest contiguous deviations across the entire search space.
Such a strategy was first deployed in experiments at the Tevatron and HERA~\cite{D0:2000vuh,H1:2004rlm,H1:2008aak,CDF:2007iou,CDF:2007ykt,CDF:2008voc} and has been continued by current LHC experiments~\cite{ATLAS:2018zdn,CMS:2020zjg}.
This approach makes minimal assumptions about the signal but assumes backgrounds are well modeled by simulation. 
While this method has the benefit of covering a large signal parameter space, it has several limitations.
Anomalies are again generally assumed to appear as contiguous deviations in the selected observables.
Moreover, since no multivariate information is utilized, anomalies appearing in final states with large backgrounds are likely to be missed.
Finally, the heavy reliance on simulation, which may have imperfections in exotic final states, limits the sensitivity of the search.
Modern techniques seek to improve upon this scenario in one or more respects.

In the last several years, there has been a significant body of work on new classes of model-agnostic search strategies enabled by advances in machine learning \cite{LHCO,darkmachines}. 
Excitingly, these new methods have now begun to be adopted by the large experimental collaborations as a complementary component of their search programs \cite{ATLAS_cwola,ATLAS_two_body, ATLAS_Higgs_anomaly, ATLAS_anomaly, ATLAS_semivisible_anomaly, ATLAS_multilep_anomaly, CMS_CASE, CMS_CASE_MLG, CMS_Higgs_anomaly}.
Though new physics has not yet been found, these searches have demonstrated the power of these approaches to discover new phenomena that may have been missed by conventional model-dependent approaches. 
It is likely that in the coming years the usage of these methods in collider searches will continue to grow, and that their potential will increasingly be recognized in other areas of physics, such as cosmology and astrophysics, where similarly complex datasets and limited theoretical guidance motivate the use of flexible, model-agnostic approaches.

While these searches have great promise, the new methods they employ and their underlying philosophies are unfamiliar to many researchers in the physical sciences.
Standard practices on the validation of these methods, and how results from these searches should be reported, have yet to be established.

In this document we attempt to close this knowledge gap, providing a concise review of the methods and strategies for their validation. 
In Section \ref{sec:methods}, we overview the basic statistical formalism of searches, and then review the major classes of new model-agnostic techniques, focusing on the major conceptual points rather than machine learning specifics.
In Sections \ref{sec:nplm} and \ref{sec:dijet_resonance} we focus on methods for the validation of these strategies in two case studies. 
In Section \ref{sec:interp}, we further discuss interpretation strategies for these new methods, both for interpreting a significant excess and methods to derive exclusion limits.
We conclude in Section \ref{sec:conclusion}.

This article contributes to VERaiPHY (Validation \& Evaluation for Robust AI in PHYsics), a PHYSTAT review series establishing verification and validation standards for machine learning across particle physics, astrophysics, and cosmology.

\section{Foundations of model-agnostic searches}
\label{sec:methods}

Existing model-agnostic search strategies can be broadly grouped into two categories.
The first category comprises methods that perform the entire statistical test: they take the observed data sample as input and directly output the statistical significance of any deviation from expectations.
Such approaches can be used as a standalone analysis or run in parallel with traditional model-dependent searches targeting a specific final state.
In this review we discuss these strategies within the formalism of \emph{two-sample hypothesis testing}. 
The second category, which we denote as \emph{model-agnostic signal selection strategies}, includes approaches that form only part of the full analysis, and do not include a statistical test as part of the method. These methods function as anomaly detectors, identifying interesting events or regions of phase space by assigning an anomaly score in a model-agnostic manner.
The selected anomalous events can then be further analyzed using either model-aware or model-agnostic statistical procedures to determine whether they are compatible with expectations under the reference background hypothesis.

In this section we first discuss the common statistical formalism of searches in high energy physics, and then overview these two complementary classes of model-agnostic search strategies. 

\begin{figure}[h]
    \centering
    \includegraphics[width = 0.7 \textwidth ]{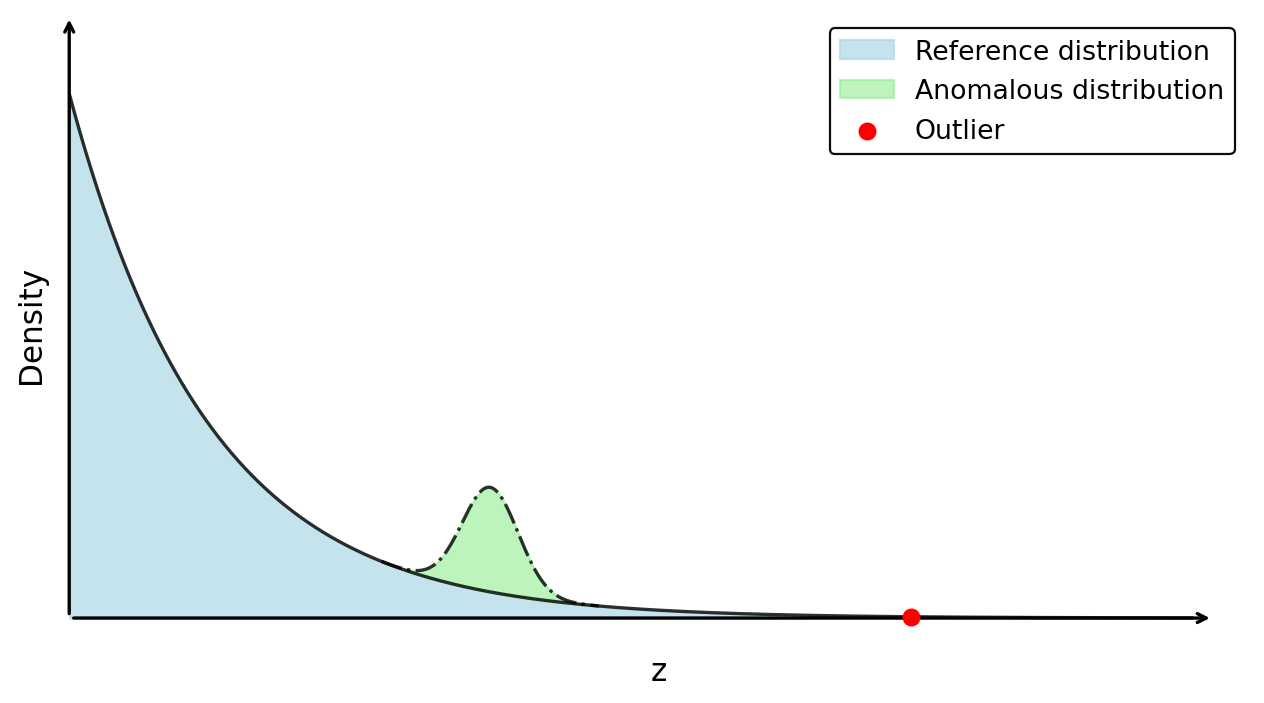}
    \caption{One-dimensional illustration of the difference between a collective anomaly, represented by a Gaussian bump on top of a reference exponential distribution, and an out-of-distribution event.}

    \label{fig:coll_outlier}
\end{figure}

\subsection{Statistical formalism of searches}\label{subsec2.1}

In the physical sciences, and particularly in fundamental physics, it is often of interest to determine whether a set of measured data deviates from the expected reference background distribution predicted by a corresponding reference background model (our current state of knowledge, for example, the Standard Model of Particle Physics or the $\Lambda$CDM model in cosmology). This can be expressed as testing whether the distribution of the measured data includes an additional signal component on top of the background distribution
\begin{equation}\label{data_distr}
    p_{\rm data}(z) = (1-\lambda)\,p_b (z) + \lambda\,p_s(z).
\end{equation}
Here $p_s(z)$ and $p_b(z)$ denote the probability distributions of the signal and the background, respectively. The signal distribution is unknown, while the background distribution is typically not available in closed analytical form but can often be sampled either through Monte Carlo simulations of the underlying physical processes or through measurements in signal-free control regions. The parameter $0 \leq \lambda \leq 1$ represents the signal strength and is also unknown \footnote{We note that this not the most general parameterization, as new phenomena can also introduce effects not parameterizable by a single linear parameter, (e.g. negative deviations from the reference via quantum interference effects), however we omit these complexities for now to illustrate the main ideas}. 

When the signal component is unspecified or unknown, the search for this type of population-level discrepancies is sometimes referred to as \emph{collective anomaly detection}. This approach stands in contrast to pointwise anomaly detection—such as outlier or out-of-distribution detection—which focuses on determining whether individual data points are atypical under $p_b$. See Figure~\ref{fig:coll_outlier} for an illustrative example.

\begin{figure}[h]
    \centering
    \includegraphics[width = 0.7 \textwidth ]{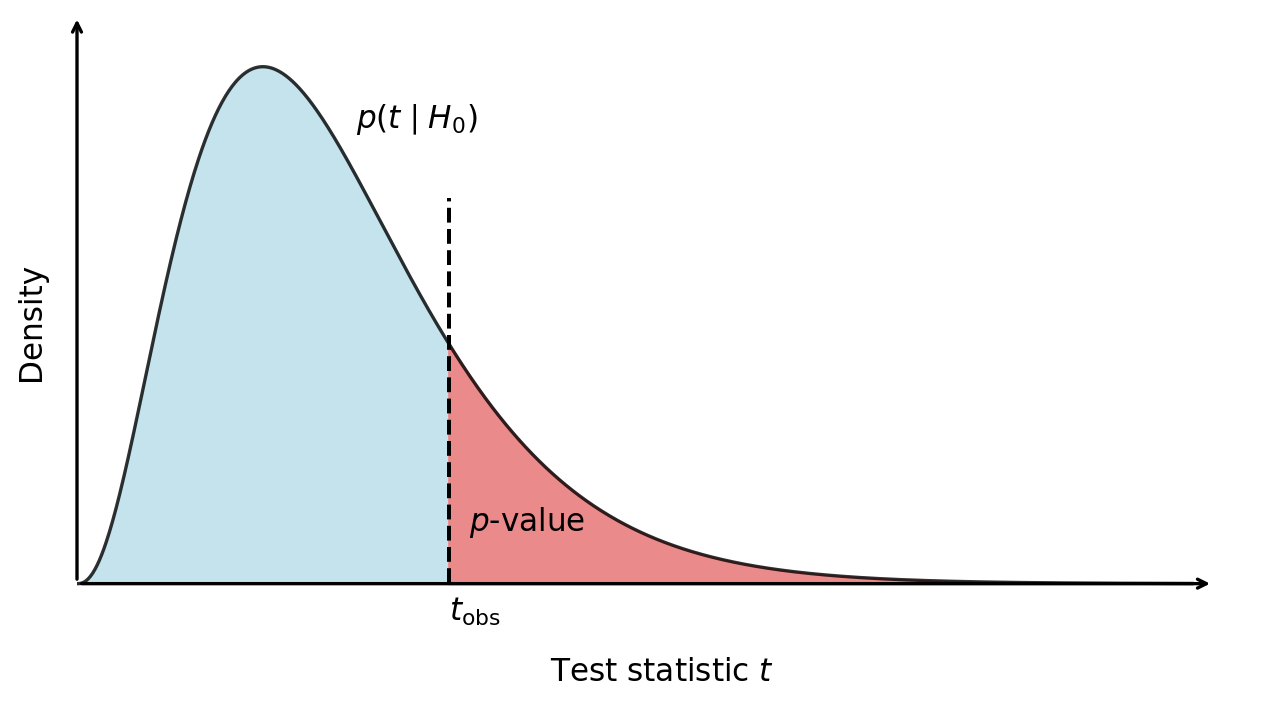}
    \caption{Illustration of the distribution of the test statistic under the null hypothesis and the $p$-value (red region).}
    \label{fig:p-val}
\end{figure}

To perform a complete statistical test for the presence of a signal, a test statistic $t$ is defined to measure how much the data differs from the background only hypothesis.
Large values of the test statistic indicates a potential tension with the null (background-only) hypothesis $H_0$.
To quantify this statement, the distribution of the test under the null hypothesis $p(t|H_0)$ needs to be known or estimated. Consequently, the $p$-value is defined as
\begin{equation}\label{eq:p-value}
\textrm{p}_{\rm value}=P(t\geq t_{\rm obs}|H_0)=\int_{t_{\rm obs}}^\infty p(t|H_0) dt,
\end{equation} 
where $t_{\rm obs}$ is the value of the observed test statistic, as illustrated in Fig.~\ref{fig:p-val}. The $\textrm{p}_{\rm value}$ is then the probability to obtain data as or more extreme as the observed ones under the null hypothesis, and the result of the test is considered statistically significant if $\textrm{p}_{\rm value}$ is smaller than a pre-selected rate of type-I errors (false-positive rate), defined as
\begin{equation}\label{eq:fpr}
\alpha = P(t\geq t_\alpha|H_0).
\end{equation}
In particle physics, it is customary to express the statistical significance of a result in terms of a Z-score, defined as $Z = \Phi^{-1}(1 - \textrm{p}_{\rm value})$, where $\Phi^{-1}$ denotes the quantile function of the standard normal distribution.

In a classical model-dependent search, where both the signal (alternative) and background hypotheses are fully specified, the Neyman–Pearson lemma states that the most powerful test statistic is the likelihood ratio between the background-only and signal + background hypotheses:

\begin{equation}\label{likelihood_ratio}
L_{s+b,b} = \frac{p_{s+b}(z)}{p_b(z)} = \frac{ (1-\alpha)p_b(z) + \alpha p_s(z)} {p_b(z)} = (1-\alpha) + \alpha \frac{p_s(z)}{p_b(z)} = (1-\alpha) + \alpha L_{s,b}.
\end{equation}

As we can see, this is just a monotonic rescaling of the likelihood ratio between signal and background $L_{s,b}$ \footnote{ Note that the monotonicity between $L_{s+b,b}$ and $L_{s,b}$ only holds for a single observation. When analyzing a collection of observations, the likelihood for the full collection will be a product of these single-observation likelihoods and due to cross terms there is no simple monotonic relationship between the $\prod_{i} L_{s,b}(x_i)$ and $\prod_{i} L_{s+b,b}(x_i)$}.
It can be shown that a supervised machine learning classifier trained to discriminate between two datasets effectively learns a monotonic transformation of the likelihood ratio. This property, often referred to as the likelihood-ratio trick~\cite{hastie2009elements}, has many useful applications in hypothesis testing. In practice, this means that a classifier trained to distinguish labeled signal and background events, for example using simulated samples, can construct a nearly optimal test statistic. However, such a methodology requires a fully specified signal hypothesis. The challenge for model-agnostic searches is therefore to design powerful test statistics without assuming a specific signal model.

\paragraph{Global and local $p$-value} In searches for new phenomena, the statistical interpretation of an observed excess is commonly expressed in terms of a \textit{local} and a \textit{global} $p$-value. These two quantities differ when multiple hypothesis tests are performed in a single search. This often occurs because some parameter of the signal such as an invariant mass is unknown and a different hypothesis test is performed for different candidate values. The local $p$-value quantifies the probability, under the null hypothesis, of obtaining a fluctuation at least as significant as the one observed at a specific point in the parameter space being tested (for instance, a particular hypothesis of the invariant mass). This measure reflects the local incompatibility of the data with the null hypothesis but does not account for the fact that many such hypotheses may have been tested. When multiple hypotheses have been tested (e.g., a search across a wide mass spectrum), the probability of observing a large fluctuation for at least one hypothesis. This so-called \textit{look-elsewhere effect} (LEE) is corrected for by evaluating the \textit{global} $p$-value, which represents the probability of obtaining, anywhere in the search region, an excess at least as significant as the one observed. As a result, the global $p$-value is typically larger than the corresponding local value, leading to a reduced global significance once the LEE is taken into account. This effect is known in the statistical literature as the \textit{multiple testing problem}. Though not directly a manifestation of the LEE, model-independent searches often face a similar tradeoff between the breadth of signal hypotheses being covered and sensitivity to detect any particular signal hypothesis. 

\paragraph{Simple and composite hypotheses} In the context of hypothesis testing, a distinction is made between \textit{simple} and \textit{composite} hypotheses. A simple hypothesis specifies the probability distribution of the data completely, with all parameters fixed (for example, a background-only model with known normalization and shape). This is the case, for instance, in fully model-dependent searches. In contrast, a composite hypothesis encompasses a family of possible distributions characterized by one or more free parameters, such as an unknown signal strength or particle mass. According to the Neyman--Pearson lemma, for testing two simple hypotheses there exists a \textit{most powerful} test, that is, a test that maximizes the probability of correctly rejecting the null hypothesis for a given significance level. However, this result does not extend to composite hypotheses: when continuous signal hypotheses are involved (such as a signal with an unknown cross section or mass), no single test statistic can be uniformly most powerful across all parameter values \cite{Carzon:2025isu}. In practice, one constructs tests based on the likelihood ratio, often using profile likelihood methods, which typically provide tests with good sensitivities and frequentist properties even in the absence of a strictly most powerful test. 

\paragraph{No optimal model-independent test} In model-independent tests, the set of alternative hypotheses may be quite large.
One would ideally design a statistical test that has maximum power to detect deviations coming from all possible alternatives. However, it has been proven that this is not possible. Statistical tests cannot have power to detect all alternative hypotheses \cite{janssen2000global}. This means that any model-independent search strategy will be insensitive to some set of deviations. 
Therefore, it behooves these strategies to use a set of physically motivated assumptions or parameter choices in their construction, so that they achieve sensitivity to genuine physical anomalies and diminish their sensitivity to unphysical deviations caused by statistical noise or instrumental failures. 
This also motivates the use of multiple strategies, which may make complementary assumptions, and thus have complementary sensitivities, in the search for unknown signals. 

\subsection{Two-sample testing for model-independent searches}\label{subsec:coll_AD}

A natural framework to formalize the statistical methodology for model-independent searches is through \emph{two-sample hypothesis testing}. Suppose we have two datasets, $\mathcal{X} = \{x_1, \dots, x_n\}$ and $\mathcal{Y} = \{y_1, \dots, y_m\}$, with points in a $d$-dimensional space, $x_i,y_j\in\mathbb{R}^d$, drawn from $p_b$ and $p_{\rm data}$ respectively. The goal of the statistical test is to asses whether the null hypothesis that both sets come from the same distribution,  $H_0: p_b = p_{\rm data}$ (hence $\alpha=0$ according to Eq.~\eqref{data_distr}), can be rejected. The alternative hypothesis $H_1$ is simply the negation of $H_0$ and no specific signal hypothesis is introduced.
In the context of searches for new physics, the null hypothesis corresponds to the background-only hypothesis, where the observed data are consistent with the predictions of the Standard Model. Conversely, the alternative hypothesis implies a deviation from the Standard Model expectation, such as the presence of a new particle or interaction. Within the landscape of Fig. \ref{fig:landspace}, these approaches make strong assumptions about the background, as it requires the reference distribution to be known well, and minimal assumptions about the signal. 

A test statistic for a two-sample test can then be defined as a function of the observed and reference data: 

\begin{equation}\label{eq:test_stat}
 t : \mathbb{R}^{n \times d} \times \mathbb{R}^{m \times d} \to \mathbb{R}.
 \end{equation}

The distribution $p(t | H_0)$ is often not available in closed analytical form and must therefore be estimated. Common approaches rely on randomized resampling methods such as permutation tests or bootstrapping. In a permutation test, for instance, the test statistic is repeatedly computed on random reshufflings of the labels characterizing the reference and the data samples, thereby generating an empirical distribution of possible outcomes under the null hypothesis \cite{wasserman2013all}. If the data contain a contribution from new physics, the observed value of the test statistic on the original partition will typically appear extreme compared to those obtained from the randomized datasets, since the new-physics contribution becomes diluted by mixing signal and background events. Alternatively, when a reliable generator of background data is available, the null distribution can be estimated by repeatedly evaluating the test statistic on pairs of independent samples drawn from the background distribution $p_b$, thereby strictly satisfying the background-only hypothesis. This is a common situation in particle physics, where, even though Standard Model predictions are well understood, complex detector effects make the probability distribution of the data analytically intractable, making the use of sophisticated Monte Carlo simulators essential.

\paragraph{The role of ML-based methods} This type of model-independent collective anomaly detection is particularly demanding in high-precision fields such as HEP. Difficulties arise from both the large number of events and the high-dimensional nature of the datasets, and the expectation that deviations from the background model may be small (exhibiting a poor signal-to-noise ratio), hidden (appearing in uncommon or weakly constrained observables), or both. Traditional two-sample test approaches are either one-dimensional (e.g., the Kolmogorov–Smirnov test) or rely on binning the data (e.g., a binned $\chi^2$ test), which becomes infeasible in more than a few dimensions and is strongly affected by the choice of binning scheme.
Machine learning-based methods for two-sample testing have been proposed in the last few years as promising approaches to address these challenges, due to their ability to fit complex patterns in multidimensional data. 
A class of proposals is based on the idea of using classifiers to separate the background data from the measured data (see for instance Refs.~\cite{friedman2003multivariate,lopez-paz2017revisiting,DAgnolo:2018cun,DAgnolo:2019vbw, kim2021classification,Letizia:2022xbe,chakravarti2023model}), without explicit hypotheses on the nature of potential anomalies. Classifier performance metrics, such as accuracy or the area under the ROC curve, can then serve as test statistics. By exploiting the ability of classifiers to learn the likelihood ratio, one can design powerful, data-driven hypothesis tests that mimic the Neyman--Pearson construction. Recent studies~\cite{chakravarti2023model,Grosso:2023scl,Grossi:2025pmm} have shown that such likelihood-ratio--based approaches can outperform traditional classifier metrics in sensitivity. Other approaches, inspired by the data-science and machine learning literature, are based on introducing a test statistic from a notion of distance between distributions that is multivariate in nature. Examples of these methods include the Kullback–Leibler divergence, the maximum mean discrepancy~\cite{JMLR:v13:gretton12a, chatalic2025efficient} and the Wasserstein distance~\cite{ramdas2017wasserstein,Grossi:2024axb,tran2025minimax}.

ML-based methods also fall within the framework of composite hypothesis testing. Classifiers operate under specific training assumptions and (hyper-)parameter choices. Since the true underlying distributions and nuisance parameters are not known exactly, the resulting test statistics are not guaranteed to be most powerful in a uniform sense. Nevertheless, they can offer near-optimal sensitivity within the region of the parameter space where the training is representative, effectively serving as flexible, data-driven approximations to the exact likelihood ratio. This challenge is closely related to the multiple testing problem discussed above: each distinct choice of model architecture, training dataset, or hyper-parameter configuration effectively defines a separate test. Recent proposals have suggested leveraging this fact to improve the sensitivity of learning-based analyses by aggregating results from multiple trained models or hyper-parameter settings \cite{biggs2023mmd, schrab2023mmd, Grosso:2024wjt}. Such ensemble-based strategies can indeed enhance discovery potential by capturing complementary features of the data. However, they do not yield a single test that is most powerful under all possible alternatives, and if applied excessively they can lead to an overall loss of power due to overfitting or implicit trials effects.

\paragraph{The reference hypothesis} One difficulty in the application of two-sample testing methods is the construction of the reference hypothesis. 
In HEP, we are fortunate to have access to very high quality simulators which can be used to simulate the standard model and encode the reference hypothesis.
For some final states these simulators are sufficient to describe the data within known systematic uncertainties and can be used to conduct a two-sample test.
However, in many other final states, particularly those involving contributions from backgrounds relating to the strong nuclear force, it is known that the simulators
are not of a sufficiently high quality to accurately describe the data with the necessary precision. 
In standard supervised analyses these final states therefore require data-driven methods to estimate the normalization and shape of the standard model background.
Whether these methods can be extended to construct a sufficiently high-quality multi-dimensional background estimate which can serve as reference distribution for modern ML-based two-sample test methods is an open research question. Indeed, if the background estimate is not sufficiently precise, the statistical test may identify a discrepancy between the background sample and the observed data even in the absence of any new-physics signal.
\\
\smallskip

A related collective anomaly detection strategy that differs from two-sample tests are tests for the violation of a symmetry in the data in an unsupervised way \cite{Lester:2021aks,Tombs:2021wae,Taylor:2023deh,Craigie:2024bhk}. 
The basic idea of these methods is to test if a symmetry-transformed version of the data is statistically distinguishable from the original data.
If so, it means the symmetry is violated in some way. 
Such symmetry violations are collective phenomena, related to distributional properties of the data rather than specific instances. 
They differ from the aforementioned general methods in that they focus on a particular type of symmetry-violating alternative hypothesis.
This limits their scope of alternatives, but allows them to be performed without an explicit reference distribution needed for a typical two-sample test.

\subsection{Methods for model-agnostic signal selection}

Rather than directly performing the full statistical test for the presence of anomalies, some methods seek instead to identify subsets of the data which are enriched in signal events. 
This is often employed when one does not have a full model of the reference hypothesis, meaning a direct two-sample test cannot be performed.
Instead, the potentially anomalous subset is identified, and then used in different ways depending on the application.
In some cases, a statistical analysis is performed on the anomalous subset, using some auxiliary information to estimate their likelihood under the null hypothesis, and obtain a $p$-value. 
In other applications, often realtime anomaly detection systems, the anomalous subsets are saved for later downstream analysis, or flagged for human inspection, but no statistical analysis is directly performed. 
Approaches to the task of model-agnostic signal-enhancement can be generally categorized into two classes.

The first class of techniques is based on the idea of \emph{outlier detection} (see Figure~\ref{fig:coll_outlier}).
Usually in the search for anomalies, background events dominate the data sample.
Portions of the data sample known to have negligible signal can be used to learn the distribution of the dominant background. 
Data instances which are very unlikely under the background distribution can therefore be considered anomalies. In essence, this technique defines $\frac{1}{p_b(z)}$ as an anomaly score. 

The second set of techniques, called \emph{weak supervision}, perform a type of collective anomaly detection to learn the unique characteristics of the signal that distinguish it from background. 
Classifiers are trained to distinguish between a subsample of the data containing potential anomalies and a data-driven estimate of the background. 
If there is an anomalous signal present in the data subsample, a classifier will learn $\frac{p_s(z)}{p_b(z)}$, the optimal signal versus background classifier.
This classifier can then be used to identify anomalous events on an orthogonal data sample.
Constructing appropriate samples for weakly supervised training relies on additional domain-specific assumptions, often leveraging the localization of the signal in some feature.

There are additional techniques which live somewhere in between full model agnostic approaches and traditional supervised methods. These usually use some representative signal models as a loose prior \cite{Park:2020pak,Cheng:2024yig}. As the focus of this review is on fully model-agnostic methods, we will not discuss them further.

\subsubsection{Outlier detection}
Outlier detection is based on learning the multidimensional distribution of background events and then identifying anomalies as events that are dissimilar with respect to this learned distribution (see Figure~\ref{fig:coll_outlier}). Since data samples are typically dominated by background processes, these methods are often trained directly on a subset of the data itself.

Learning multidimensional probability distributions that allow for direct estimation of the probability density is a challenging task, so many applications instead rely on a proxy objective to encode the background probability density. A commonly used machine-learning model for outlier detection is the autoencoder, first proposed for applications in particle physics in \cite{Heimel:2018mkt,Farina:2018fyg}. Autoencoders are neural networks that take input data of dimension $Z \sim \mathbb{R}^d$ and encode it, via an encoder network $E(z)$, into a latent space of smaller dimension $Y \sim \mathbb{R}^k$, with $k < d$. A decoder network $D(y)$ then maps this latent representation back to the original space in an attempt to reconstruct the input.
Formally, the network is defined by $E(z)=y$ and $D(y)=z'$, and it is typically trained by minimizing a L2 reconstruction loss, $\mathcal{L} = \lVert z - D(E(z)) \rVert^2$.

When trained on a sample dominated by background events, the autoencoder learns to perform this compression and decompression efficiently for such events. An anomalous event is then effectively \emph{out of distribution} with respect to the training data, causing the autoencoder to reconstruct it poorly. The resulting L2 reconstruction loss can therefore be used as an anomaly score.
This L2 loss can be seen as proxy for a quantity like $\sim \frac{1}{p_b(z)}$, but in practice it has several limitations.

An alternative strategy to autoencoders is to learn $p_b(z)$ directly through density estimation techniques.
This can be accomplished, for example, using variational autoencoders \cite{kingma2014autoencoding, Cerri:2018anq}, which enhance the original autoencoder architecture by enforcing a multivariate Gaussian structure in the latent space via additional terms in the loss function.
The Gaussian structure allows the estimation of the likelihood of a data point, by first transforming it into its latent vector $y$ and then evaluating the likelihood of $y$ under the known multivariate Gaussian distribution. 
Other machine learning models, such as normalizing flows or flow matching diffusion models\cite{rezende2016variational,lipman2024flowmatchingguidecode}, also allow multivariate density estimation and can therefore be used for outlier detection \cite{Mikuni:2023tok, Vaselli:2025zkl}.
These models are generally believed to scale more effectively and to model complex multivariate densities more accurately than variational autoencoders. 

\paragraph{Fundamental Limitations}
One inherent limitation of all outlier detection methods, which implicitly define anomalies as regions of low probability densities, is that probability densities are not invariant under coordinate transformations (see the work in Ref.~\cite{Kasieczka:2022naq}, product of a discussion that took place at the \href{https://doi.org/https://indico.cern.ch/event/1138933/}{PhyStat-Anomaly workshop}).
This means that the notion of regions of low probability density, and therefore what an outlier detection method defines as an anomaly, depends on the coordinate system. 
Under an invertible transformation of the data $y = f(z)$ , the probability density changes to

\begin{equation}
    p_y(y) = p_z(f^{-1}(y)) | \frac{d}{dy} f^{-1}(y)|
\end{equation}

where the last term is the Jacobian of the transformation. 
For non-trivial mappings, this Jacobian can radically alter the location of high- and low-density regions when going from $x$ to $y$. 
or example, if $y=z^2$, and $p_y(y) \sim e^{-ky}$ then $p_z(z) \sim z e^{-kz^2}$. 
This change of variables radically alters the interpretation of the $y=z=0$ point, as $p_z(0)$ is the peak of the probability distribution but is the minimum for $p_y(0)=0$.
Alternatively, if $f(x)$ is the cumulative distribution function of $x$, then in $y$ all points will have uniform density and no point will be rarer than any other.
In practice this means that the choice of data representation and pre-processing transformations define significant inductive biases that determine what kind of anomalies the method will be sensitive to.
Signals which do not live in the low-density regions of the chosen data representation will be missed by outlier detection methods.
Therefore, significant care should be put in the choice of data representation for any outlier detection strategy.
See Ref. \cite{OOD_answer} for an extended discussion of these limitations.

Note that for ratios of probability densities, like $L_{s,b}(x)$, the Jacobian of the coordinate transformation in the numerator and denominator cancels out and therefore the classification score is invariant. This is one of the main advantages of likelihood-ratio-based methods, as discussed in the other sections.

Another significant challenge specific to outlier detection-based methods is their so called \emph{complexity bias}.
For autoencoders, because the anomaly score is based on a compression task, more complex data instances (of a higher intrinsic dimension) tend to receive larger anomaly scores regardless of whether they are present in the training sample. 
Interestingly, similar biases have been observed for density estimation methods when evaluating out of distribution samples~\cite{Gordon_OOD,serra_OOD}.
One manifestation of this bias is that an autoencoder trained exclusively on QCD jets is able to identify top jets as anomalous, whereas an autoencoder trained on top jets struggles to identify QCD jets as anomalous \cite{Buss:2022lxw}.
Normalized autoencoders \cite{Dillon:2022mkq, CMS_WNAE} attempt to mitigate this issue by turning autoencoders into energy based probabilistic models. 
In this approach, the network is penalized for accurately reconstructing out-of-distribution data, leading to a better representation of $p_b(z)$ than that obtained with a standard autoencoder.

\subsubsection{Weak supervision}

Weak supervision seeks to train a classifier to learn to identify anomalies using only the data sample.
Typical classifiers rely on labeled events for training, which are not available for most data samples containing anomalies. 
In Classification Without Labels (CWoLA)~\cite{Metodiev:2017vrx}, noisy labels based on mixed samples of events are used instead.
Suppose one has two samples, $M_1$ and $M_2$, which are composed of a mixture of signal and background events.
The composition of each sample is unknown, but for some reason one knows that $M_1$ has a larger fraction of signal events in it ($f_1$) than $M_2$ does ($f_2$).
Then, training a classifier to distinguish between from $M_1$ and $M_2$ will converge to the optimal signal versus vs background classifier.
This is because the likelihood ratio between $M_1$ and $M_2$ is just a rescaled version of the likelihood ratio between signal and background:

\begin{equation}
    L_{M1,M2}(z) = \frac{p_{M1}(z)}{p_{M2}(z)} = \frac{f_1 p_s(z) + (1-f_1) p_b(z)}{f_2 p_s(X) + (1-f_2) p_b(z)} 
    = \frac{f_1 L_{s,b} + (1 - f_1)}{f_2 L_{s,b} + (1 - f_2)}.
\end{equation}

One can check that for $f_1 > f_2$ this is just a monotonic rescaling of $L_{s,b}$ and therefore defines an equivalent classifier. 
This is clear in the limit of $f_2 \to 0 $, which occurs when the $M_2$ sample is essentially pure background, which occurs in many HEP applications of this technique.

\begin{figure}
    \centering
    \includegraphics[width = 0.5 \textwidth ]{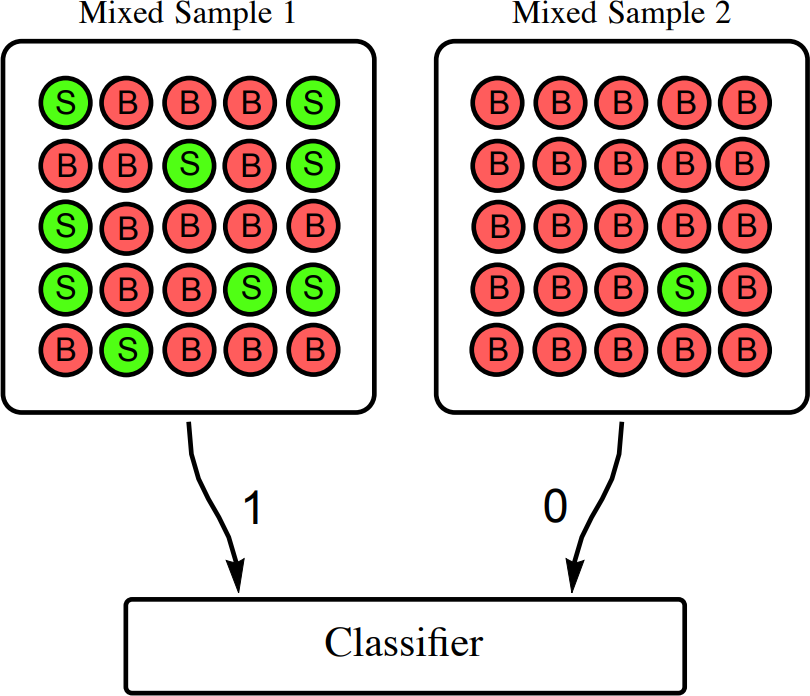}
    \caption{An illustration of weakly supervised training. A classifier is trained to distinguish between two mixed samples of signal and background events. Taken from \cite{Metodiev:2017vrx}. }
    \label{fig:weak_sup}
\end{figure}

The training setup shown graphically in Figure \ref{fig:weak_sup}.

The key assumption underlying weak supervision is that the background events in the two samples are sampled from the same underlying distribution. 
If this is true, the only way to distinguish the two samples is the difference in relative signal fractions between the two samples so the classifier will learn to distinguish signal versus background. 
If there is any bias such that the background events from the two samples do not come from the same distribution, then this will typically dominate the loss (because anomalies are typically rare, so $f_1 << 1$) such that the network will learn this background bias rather than signal vs background discrimination. 

To apply this technique to anomaly detection, one must define a method to construct the mixed samples $M_1$ and $M_2$ from the unlabeled data. 
There is no generalized procedure to do this. 
Applications of weak supervision rely on domain-specific physics knowledge to appropriately define the samples.
As discussed further in Section~\ref{sec:dijet_resonance}, many different techniques have been proposed to construct the $M_2$ sample based on interpolation, reweighting and transport methods \cite{Collins:2018epr,Collins:2019jip, Amram:2020ykb, ANODE, Hallin:2021wme, Hallin:2022eoq, Stein:2020rou, Andreassen:2020nkr, Benkendorfer:2020gek, Raine:2022hht, Golling:2022nkl, Kamenik:2022qxs, Chen:2022suv,Sengupta:2023xqy, Finke:2022lsu,Bickendorf:2023nej, Golling:2023yjq, Buhmann:2023acn, SIGMA, RANODE, Bai:2023yyy, Kasieczka:2024lxf}.

The most well studied application of weak supervision is for resonant searches, first proposed in \cite{Collins:2018epr,Collins:2019jip}.
In a weakly supervised resonance search the signal is assumed to be localized in a narrow region of some pre-defined resonance mass. This allows the signal-enriched sample, $M_1$ to be defined using a window in the resonant variable, and the $M_2$ sample can be constructed through interpolation of background events outside this window.
An example application of this method to a resonant signal is given in Section~\ref{sec:dijet_resonance}.

Similar methods have also been applied to astrophysical data to automate the detection of stellar streams \cite{Shih:2021kbt,shih2023machinae,pettee2023weaklysupervised,sengupta2024skycurtains}.
Exploring additional domains where such samples can be constructed and weak supervision applied is an open research direction.

Because the weakly supervised classifier is trained on events from the signal region of the analysis, it should then not be applied to those same events to identify anomalous events.
Otherwise, an overfitting of the classifier would lead to a bias in the search result.
However, one would like to avoid 'wasting' some portion of the data to only to train the classifier, which would reduce the statistical sensitivity 
of the search.
Instead, $k$-fold cross-validation schemes have been used, in which the data sample is split into $k$ separate folds. 
The data from $k-1$ folds are used to train the classifier which is then applied to the $k$th fold to select events and perform statistical analysis. 
The procedure is then repeated $k$ times, rotating usage the different folds so that each fold is used to select events exactly once. 

It has recently been pointed out \cite{WeakSup_LEE, look_everywhere} that because a given region is used to both define the event selection and search for an excess, statistical 
fluctuations can be amplified, incurring an effective LEE.
This means that the $p$-values from any statistical analysis making use of cross-validation should be a considered a `local' value which must be calibrated with toys
to determine its global significance. 
More research needs to be done to establish best practices to mitigate this affect and properly calibrate global $p$-values in a computationally tractable manner.

A well-understood statistical effect in weakly supervised searches is that the efficiency to select signal events as anomalous depends strongly on the amount of
signal in the dataset. 
If there is no signal present in the dataset, then the weakly supervised training procedure will have the impossible task of attempting to differentiate two datasets of pure background events.
The resulting classifier will then likely overfit some random statistical difference between the two samples.
It will therefore have very low efficiency at selecting any anomalies.
However, if there is a large amount of signal present in the dataset, the asymptotic properties of weak supervision discussed above will manifest, and the performance classifier will approach that of a supervised classifier, resulting in large signal efficiency. 
For intermediate signal strengths the performance increases as a function of the signal strength.
As discussed in Section \ref{subsec:limits}, this property complicates the extraction of exclusion limits from weakly supervised anomaly detection searches.  

\subsection{Comparison of Approaches}

\begin{figure}
    \centering
    \includegraphics[width=0.9\linewidth]{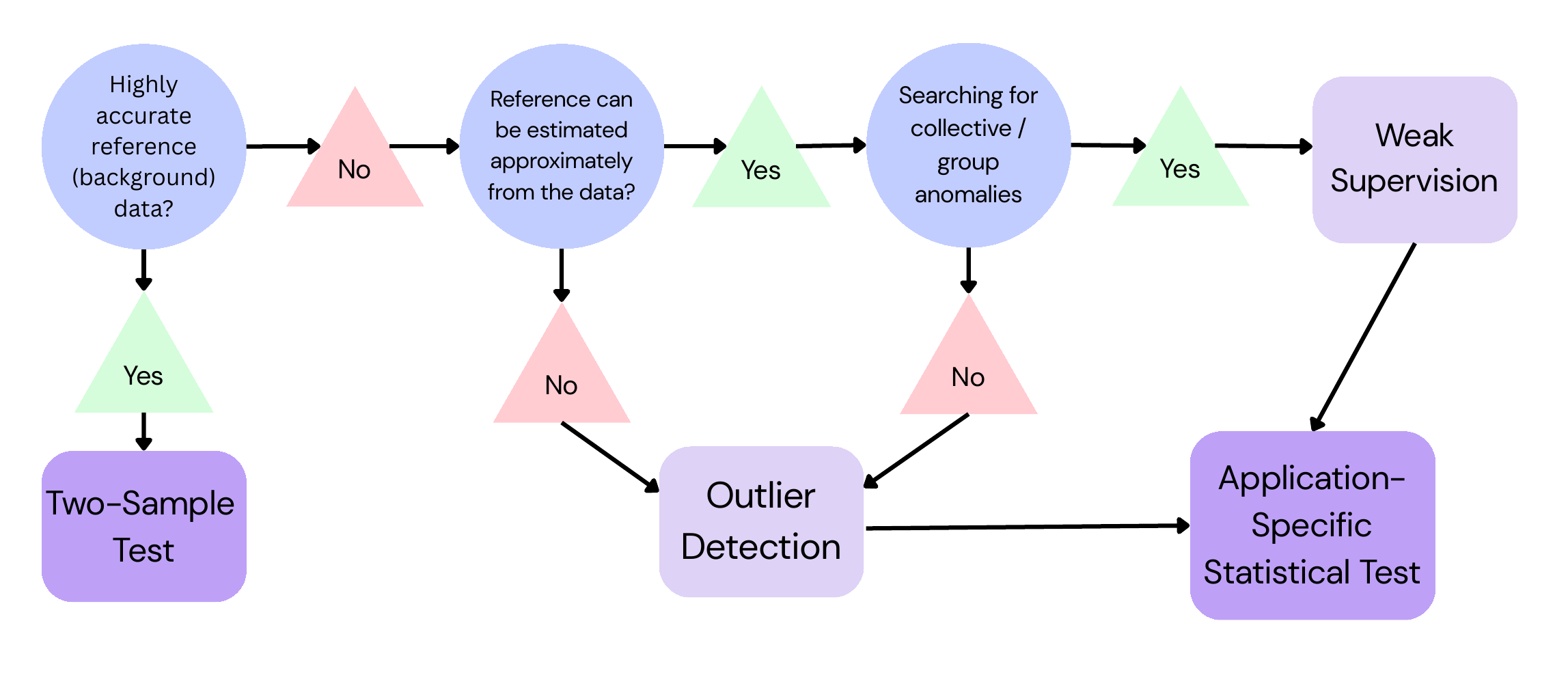}
    \caption{A flowchart illustrating an proposed set of criteria to determine when to use the three main classes of anomaly detection methods.}
    \label{fig:AD_flowchart}
\end{figure}

Given these various options, one might wonder which method should be employed in a given application. We provide a recommendation in terms of a simplified flowchart in Fig. \ref{fig:AD_flowchart}. 

When one has access to high quality reference data encoding the null hypothesis, we recommend two-sample tests as they are arguably the most signal-model-independent strategy. 
Two-sample tests also perform a full statistical test whereas the other methods require an additional application-specific strategy to extract a statistically meaningful statement. 
However, in many cases one does not have the required high quality reference data, in which case one of the signal-enhancing methods must be employed.
In this scenario, weak supervision is preferred when searching for group / collective anomalies, due to their coordinate invariance and asymptotic optimality properties. 
Weak supervision requires an approximate background sample to be constructed, often in a data-driven way.
These data-driven background estimates usually necessitate some assumption on the signal (e.g. a resonance).
In situations where this is not possible, or one is not searching for a collective anomaly (such as in realtime detection applications), we recommend outlier detection strategies.
We also comment that because outlier detection strategies do not train on the signal region data they bear the most similarity to traditional search strategies and therefore may offer best ease-of-use. 

Once the signal-enhancing methods have been employed to identify potential anomalies, a statistical test still needs to be deployed to extract a significance.
Such a statistical test will require an estimate of the background, which will require the use of some domain-specific assumptions or methods (e.g. a bump-hunt fit to search for a resonance). 

It should be noted that if one would like to perform a two-sample test and retain sensitivity to outliers, an appropriate test must be chosen.
For example in the one dimensional setting, the Kolmogorov-Smirnov test is known to be insensitive to tail effects and would likely miss outliers, whereas an Anderson-Darling test might be sensitive to such outliers.
Likelihood-ratio-based tests should in principle be sensitive to such outliers. 
As if the reference density indeed goes to zero for the outlier point, the likelihood ratio should go to infinity.
However, in practice these tests rely on learning the likelihood ratio empirically from the data and may struggle to learn it well from only a single outlying data instance.
More work is needed to understand the behavior of these multivariate two-sample tests in this regime. 

\section{Two-sample Test Case Study: NPLM}\label{sec:nplm}
\subsection{Foundations}\label{intro_nplm}

The New Physics Learning Machine (NPLM) is an approach to signal-agnostic searches designed to perform a Goodness-of-Fit (GoF) test, i.e. a particular type of hypothesis test that assesses whether observed data are compatible with a given reference distribution without relying on a specific alternative hypothesis. Its purpose is to detect generic deviations from the reference model.

The Neyman–Pearson (NP) framework for hypothesis testing~\cite{Neyman:1933wgr}, by contrast, is based on comparing the relative likelihood of two competing hypotheses and provides the optimal test statistic for simple hypotheses. NPLM leverages this principle to implement a GoF test by learning an alternative hypothesis directly from the data. In this way, it combines the hypothesis-testing foundation of the NP approach with the model-independence characteristic of GoF tests.  The connection between goodness-of-fit tests and the Neyman–Pearson construction underlying NPLM was first discussed in Ref.~\cite{Baker:1983tu} and, more recently, in Ref.~\cite{Grosso:2023scl}.

We consider here NPLM as a two-sample testing case study for two main reasons. First, it provides a representative example of a broader class of model-agnostic methods based on two-sample testing, in which a flexible model is trained to distinguish observed data from a reference sample (see, e.g., Ref.~\cite{chakravarti2023model}). Second, it is currently the only approach of this type that incorporates the treatment of systematic uncertainties, which are discussed in more detail in Section~\ref{subsec:NPLM_syst}.

More concretely, the goal is to compare a reference background model $R$ (for example the SM or the $\Lambda$CDM model) with data by exploring a parametrized family of models $H_w$, which defines a composite alternative hypothesis. The method is designed to approximate the maximum log-likelihood ratio
\begin{equation}\label{lik_ratio}
t(\mathcal{X})=2\max_w\log\frac{\mathcal{L}(H_w|\mathcal{X})}{\mathcal{L}(R|\mathcal{X})},
\end{equation}
computed on the data of interest $\mathcal{X}$.
Concretely, the alternative hypothesis is defined as a local deformation of the background distribution
\begin{equation}\label{nplm_defo}
    n_w(z)=e^{f_{w}(z)}\,n_b(z),
\end{equation}
where $\mathcal{F}=\{f_w\}$ is a rich family of functions parametrised by $w$, for example neural networks or kernel methods. Here, the symbol $n(z)$ denotes a \emph{number density}, namely the probability density function normalized to the number of expected events under a certain physical hypothesis. For example, for the background model it would read $n_b(z)=N(b)\,p_b(z)$. This captures both changes in the shape of the distribution and shifts in the overall event rate, as in counting experiments \cite{DAgnolo:2018cun, Letizia:2022xbe}. In practice, as anticipated in Section~\ref{subsec:coll_AD}, a classifier is trained on measurements and background data to directly approximate the ratio of the data-generating distributions
\begin{equation}\label{eq:reweighting}
 f_{\hat{w}}(z)\approx \log \frac{n_{\rm data}(z)}{n_b(z)},
\end{equation}
where $\hat{w}$ are the optimal parameters at the end of training.

The algorithm is trained to minimize a loss function consisting of two components: a fitting term, designed to enforce Eq.~\eqref{eq:reweighting} (for example, a binary cross-entropy loss as in Ref.~\cite{Letizia:2022xbe}), and a regularization term that constrains the model’s complexity (such as an $L^2$ penalty).
At the end of training, the model is evaluated in-sample on the entire dataset using the metric
\begin{equation}\label{eq:nplm_statistic}
t_{\rm obs}(\mathcal{X},\mathcal{Y}) = -2\left[\frac{N(b)}{m}\sum_{z\in\mathcal{Y}}\left(e^{f_{\hat{w}}(z)}-1\right) - \sum_{z\in\mathcal{X}} f_{\hat{w}}(z)\right],
\end{equation}
which is a Monte Carlo–based rewriting of the extended log-likelihood ratio, as detailed in Refs.~\cite{DAgnolo:2018cun,Letizia:2022xbe}. This quantity defines the NPLM test statistic. Here, $\mathcal{X}$ denotes the data sample of interest of size $n$, $\mathcal{Y}$ a background sample of size $m$ (also referred to as \emph{the reference sample}), and $N(b)$ the expected number of measured events under the reference background model.

\begin{figure}[ht]
    \centering
    \includegraphics[width = 0.7 \textwidth ]{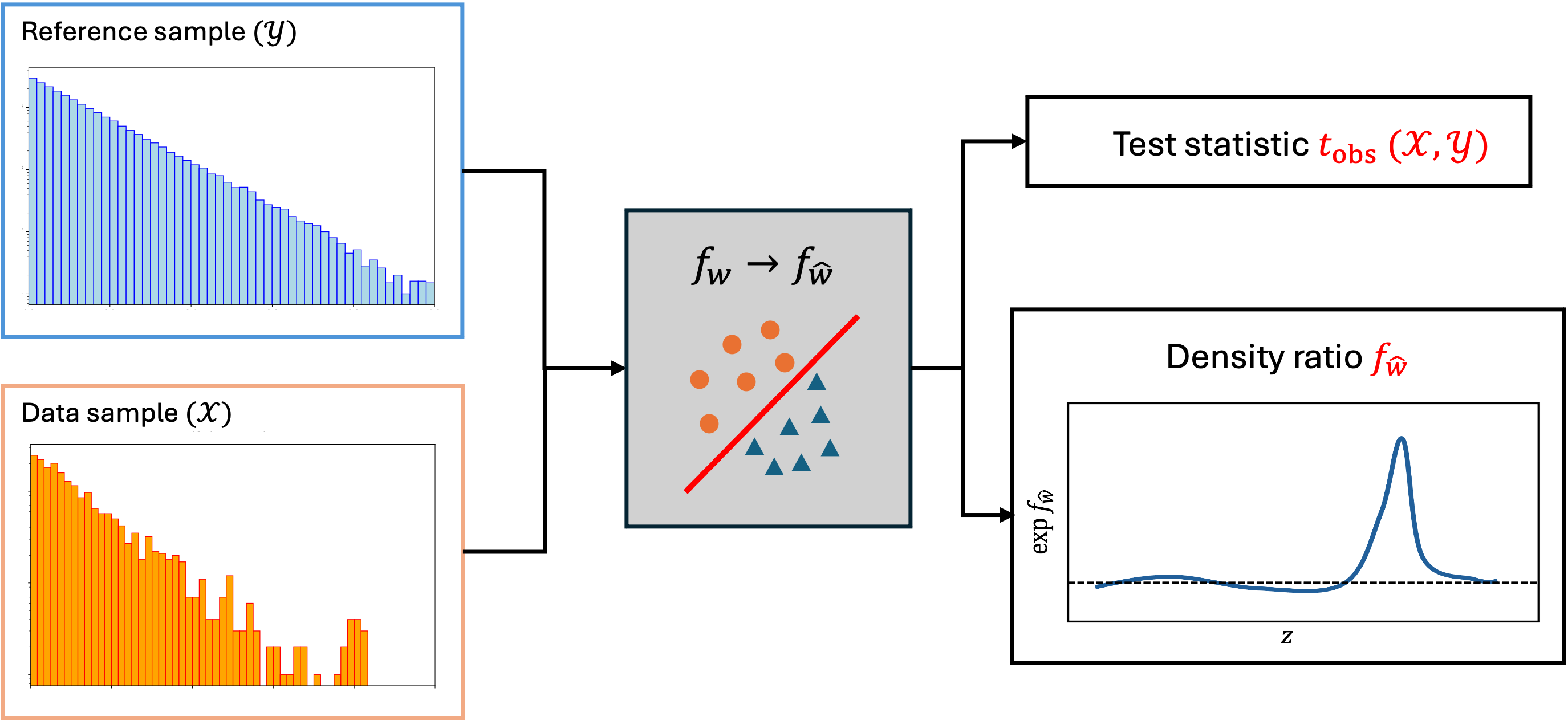}
    \caption{An illustration of the NPLM method (input data is unbinned).}
    \label{fig:NPLM_scheme}
\end{figure}

This method enables the construction of a likelihood-ratio test without the need to specify the hypotheses a priori, as they are inferred directly from the training dataset. If the data sample exhibits anomalous behavior relative to the background sample, the learned function $f_{\hat{w}}(z)$, which encodes the reweighting between the data-generating PDFs as expressed in Eq.~\eqref{eq:reweighting}, can be examined to identify the most discrepant regions in the input-feature space and their combinations, for example an invariant mass not given as an input to the model. An illustration of the pipeline is provided in Fig.~\ref{fig:NPLM_scheme}. 

The null hypothesis for the NPLM test is typically estimated through repeated evaluations of the test statistic on pairs of samples drawn from the background distribution, as described in Section~\ref{subsec:coll_AD}. In practice, at each evaluation, a reference sample $\mathcal{Y}$ is compared with a toy data sample drawn from the same background distribution to simulate measured data that are free of new-physics components. It is generally advantageous to perform the test on unbalanced datasets, with the reference sample larger than the data sample, i.e., $m > n$. This allows the model to learn an accurate representation of the reference distribution in Eq.~\eqref{eq:reweighting} and makes the outcome of the test less sensitive to statistical fluctuations affecting the reference sample.

 Ref.~\cite{Grosso:2023scl} provides comparisons with standard metrics and methods widely used in statistics and machine learning, such as the binned $\chi^2$ test, the Kolmogorov–Smirnov test, the area under the ROC curve, and classifier two-sample tests \cite{lopez-paz2017revisiting}. A recent comparison of NPLM with other statistical tests can be found in Ref.~\cite{Grossi:2025pmm}.

\subsection{Systematic uncertainties}\label{subsec:NPLM_syst}
In the context of two-sample testing for signal-agnostic searches, NPLM is currently the only approach that incorporates a treatment of systematic uncertainties affecting the simulation of background data~\cite{dAgnolo:2021aun}. The goal of this development is to enhance robustness against a potentially misspecified background model, as depicted in Fig.~\ref{fig:landspace}. The methodology is inspired by the profile likelihood-ratio approach commonly used in statistical analyses at the LHC~\cite{ParticleDataGroup:2024cfk}. Each source of uncertainty in the background Monte Carlo simulation is associated with a nuisance parameter $\nu$, so that the reference background model is promoted to a family of models $R_{\nu}$ and interpreted as a composite hypothesis. The alternative hypothesis is again formulated as a deformation of the background model, as in Eq.~\eqref{eq:test_stat}, hence depending on both $w$ and $\nu$. The test statistic the model aims at computing is now
\begin{equation}
    t(\mathcal{X})=2\log\frac{\max_{w,\nu}\mathcal{L}(H_{w,\nu}|\mathcal{X})}{\max_\nu\mathcal{L}(R_\nu|\mathcal{X})}.
\end{equation}
The main additional step with respect to the standard NPLM pipeline consists in learning how the reference background distribution deforms under variations of the nuisance parameters. This is achieved by using neural networks to approximate the density ratio between different realizations of the background model. Specifically, a classifier is trained to distinguish background samples generated with different values of the nuisance parameters from a nominal (\emph{central-value}) reference sample, thereby learning the response of the background distribution to systematic variations. Assuming that systematic effects are small, this dependence is modeled using a low-order Taylor expansion in the nuisance parameters,
\begin{equation}
\label{syst_ratio}
r(z;\nu)=\frac{n_{b_\nu}(z)}{n_{b_0}(z)}\approx \exp\left[\nu\,\delta_1(z)+\frac{1}{2}\nu^2\,\delta_2(z) + \cdots\right],
\end{equation}
where the functions $\delta_i(z)$ are represented by neural networks and with the series truncated at some finite order. Once the nuisance-parameter dependence has been learned, it is incorporated into the two-sample test with minimal conceptual differences with respect to the standard NPLM pipeline at the level of the test construction. The comparison between data and simulation is then performed while allowing the nuisance parameters to vary, selecting the background model that best describes the data in the absence of new physics through a profiled likelihood-ratio construction, as detailed in Ref.~\cite{dAgnolo:2021aun}. Potential discrepancies are therefore assessed relative to an optimally adjusted reference hypothesis, reducing the risk of false discoveries driven by systematic mismodeling. At the same time, the test remains sensitive to genuine discrepancies that cannot be absorbed by nuisance variations.

\subsection{The role of model selection}
With the rise of ML-powered approaches to data analysis and anomaly detection, various methods have been proposed over the past few years (the reader can find an exhaustive review in~\cite{Belis:2023mqs}), some of which have already been applied to experimental data (see for example~\cite{CMS:2024nsz} and~\cite{ATLAS:2020iwa}). Despite their potential, the adoption of these techniques introduces new challenges, particularly in understanding how model selection (the choice of hyperparameters)  can impact sensitivity and introduce biases.  
Let us consider, as an illustrative example, the case of the kernel-based NPLM test. The space of functions that is explored by the classifier to fit the density ratio is parametrized as a combination of Gaussian kernels
\begin{equation}
    f_w(z)=\sum_i w_i\, k(z,z_i),\quad k(z,z')=\exp\left[-\frac{(z-z')^2}{2\sigma^2}\right].
\end{equation}
If the bandwidth $\sigma$ (a hyperparameter) is small, the model would favor narrow resonances while, if it is large, the model will more easily detect broader excesses in the data with respect to the reference predictions. On the one hand, this can be exploited to enhance sensitivity to specific signal hypotheses of interest. On the other hand, this effect is present for any hyperparameter of a learning model, including the architecture of a neural network and the parameters driving regularization, whose impact on the outcome of the test is more opaque. If the goal is to maintain signal agnosticity as much as possible, strategies must be developed to address this issue. In~\cite{Grosso:2024wjt}, the authors explored the possibility to leverage multiple testing to ``turn a bug into a feature'', namely to combine multiple tests characterized by different choices of hyperparameters in ways that are robust against the LEE. It was shown that the sensitivity of the resulting test is more homogeneous across a number of benchmark of possible new physics signatures. This strategy can be applied to different framework beyond NPLM. However, it should be kept in mind that there is, generally speaking, a tradeoff between sensitivity and model-agnosticity.

\subsection{Validation of the null hypothesis}

In two-sample testing, the null hypothesis \(H_0\) asserts that the two data-generating distributions are identical. Informally, \emph{validation of the null hypothesis} refers to verifying that the test behaves as intended when $H_0$ is true. In practice, this means confirming that the test controls the Type~I error at the nominal level (the chosen significance level $\alpha$) and that the resulting $p$-values are uniformly distributed under $H_0$.

When the null distribution of the test statistic is estimated via permutations, and all test hyperparameters are fixed \emph{a priori} and are not selected using label-dependent procedures, these properties are guaranteed by construction. In particular, for permutation-based tests, there exists a \emph{non-asymptotic} guarantee~\cite{wasserman2013all}: if the data are exchangeable under $H_0$ and the $p$-value is computed correctly, then the test controls the Type~I error at level $\alpha$ for any finite sample size, up to Monte Carlo error due to a finite number of permutations.

Nevertheless, checking Type~I errors and the distribution of $p$-values remains good practice regardless on how $p(t|H_0)$ has been estimated, as such checks may help identify implementation errors or other bugs in the testing pipeline that are not apparent from theoretical considerations alone. Empirical validation becomes crucial when the null distribution is derived from asymptotic approximations, when multiple tests are combined, or when the data exhibit potential dependence\footnote{For example due to temporal correlations in time-series data or other forms of dependence between observations that violate the independence assumptions of the test.}, as these situations can lead to miscalibration and incorrect Type~I error rates.

NPLM is inspired by the maximum likelihood-ratio test, and it is therefore natural to investigate whether the distribution of its test statistic under the null hypothesis follows, or can be approximated by, a $\chi^2$ distribution with a number of degrees of freedom (dof) related to the number of trainable parameters $w$ in the learning model, at least in certain regimes. If such an approximation holds, it could be used to compute a $p$-value without the need to estimate $p(t | H_0)$ using toy experiments, or to ensure that the test statistic is well behaved and not heavy-tailed. This is a desirable property, as heavy tails can inflate Type~I errors or reduce statistical power.

For a standard likelihood-ratio test with composite nested hypotheses, the asymptotic distribution of $p(t | H_0)$ follows a $\chi^2$ distribution with a number of dof equal to the difference in the number of parameters between the alternative and reference models. This result does not generally hold for NPLM.  Although no formal results establish this connection, several studies \cite{DAgnolo:2018cun,DAgnolo:2019vbw,dAgnolo:2021aun,Letizia:2022xbe} suggest that regularization of the underlying learning model can lead to a regime in which approximate compatibility with a $\chi^2$ distribution is empirically recovered.\footnote{It is natural to expect a connection between the level of regularization and the amount of available data, given that asymptotic results formally apply only in the limit of infinite statistics.} This approximate condition can then be exploited to tune the model hyperparameters and obtain a well-calibrated test. The exact number of dof of the target $\chi^2$ distribution depends on the specific NPLM implementation, but is generally related to the complexity of the function space spanned by the learning model. For example, in the neural-network-based NPLM implementations of Refs.~\cite{DAgnolo:2018cun,DAgnolo:2019vbw}, it is simply given by the number of trainable parameters of the network, interpreted as the parameters characterizing the alternative hypothesis. It is worth noting that highly regularized models tend to produce test-statistic distributions that are more sharply peaked near zero. While this behavior helps avoid heavy tails, it also significantly restricts the model’s flexibility and may reduce its ability to detect sharp or highly localized anomalous features in the data.

Since current evidence for the $\chi^2$ approximation of $p(t \mid H_0)$ in NPLM is empirical rather than theoretical, $p$-values are in practice estimated using toy experiments. The $\chi^2$ fit is therefore used primarily as a diagnostic and calibration tool, for instance to guide hyperparameter tuning.

Finally, in the presence of systematic uncertainties, an additional validation step is required: one must verify that the distribution of the test statistic under the null hypothesis is approximately independent of the nuisance parameters. In practice, this involves estimating the null distribution using toy datasets generated from the reference model at different points in the nuisance-parameter space. Failure of this condition may lead to miscalibration and incorrect Type~I error rates.

\subsection{Assessing performance}
A central issue in deploying a signal-agnostic test concerns the validation of its performance. One possibility is to establish the sensitivity of the method in controlled benchmark scenarios, where the anomalous signal is specified. To obtain a robust estimate, one would ideally compare a signal-agnostic method against a well-defined ground truth. This role is naturally played by a likelihood-ratio test with fully specified signal and background hypotheses, as guaranteed by the Neyman--Pearson lemma. Since analytical PDFs are often unavailable, this test can be implemented either through standard techniques based on the choice of optimal observables and cuts, or by training a fully supervised classifier on (simulated) signal and background data to estimate the likelihood ratio, as discussed in Section~\ref{subsec2.1}.
This signal-aware test yields an estimate of the highest significance (lowest $p$-value) that any analysis can in principle attain. Consequently, the $p$-value obtained with a signal-agnostic method can be compared to this estimated \emph{ideal $p$-value}. While such comparisons are inherently dependent on the choice of benchmark signals and cannot fully characterize performance across the entire space of possible new-physics scenarios, they nevertheless provide valuable insight into the potential loss of statistical power of the method for specific classes of signals. In this sense, benchmark studies offer a controlled way to quantify how much sensitivity is sacrificed in exchange for signal agnosticity. For instance, the authors of Refs.~\cite{DAgnolo:2018cun,DAgnolo:2019vbw} used this procedure to assess the performance of the NPLM method across a limited set of HEP scenarios, thereby quantifying the typical sensitivity loss with respect to fully supervised analyses.

\subsection{Applications and prospects}
The primary motivation for deploying the NPLM method is the search for new physics at the LHC. A practical strategy in this context is to focus on a specific final-state topology and perform the analysis using a set of variables that fully characterizes the event kinematics. At its current stage of development, NPLM is best suited to studies involving a relatively small number of features, such as dimuon final states, which provide particularly clean experimental channels with systematic uncertainties that are well understood and under control. While efforts are ongoing to deploy NPLM in real data analyses within LHC physics and beyond, several studies have explored possible extensions aimed at overcoming some of its current limitations. One such approach investigates the use of \emph{foundation models} to improve sensitivity in high-dimensional settings~\cite{Metzger:2025ecl}.

More broadly, two-sample testing methods can be applied to any scenario in which one seeks to assess whether two datasets are drawn from the same underlying probability distribution. Various tasks can be naturally formulated within this statistical framework. For instance, Ref.~\cite{Grosso:2023ltd} explores the use of the NPLM pipeline for \emph{data quality monitoring}, namely the real-time monitoring of particle detectors. In this application, the model hyperparameters are chosen to prioritize fast execution over model complexity. Finally, the evaluation and comparison of data-generating methods, whether based on traditional Monte Carlo techniques or on modern generative models, can also be naturally addressed within this framework, as shown for example in Refs.~\cite{Grossi:2025pmm,Cappelli:2025myc}.

\section{Outlier Detection and Weak Supervision Case Study: Dijet Resonance Searches}
\label{sec:dijet_resonance}

The most well-studied case of classification-based anomaly detection in particle physics are resonance searches.
In a resonance search, signals manifest as a relatively narrow peak in some invariant mass distribution on top of a falling background distribution. 
Bump hunt searches have been performed for a long time in particle physics.
They make minimal assumptions and typically have sensitivity to many models of new particles.
However in final states with large backgrounds, they can loose sensitivity to rare signals as the signal bump is buried under the large background. 
Anomaly detection techniques can thus be used to enhance sensitivity to signals which produce distinctive features in addition the resonance. 

\begin{figure}
    \centering
    \includegraphics[width = 0.5 \textwidth ]{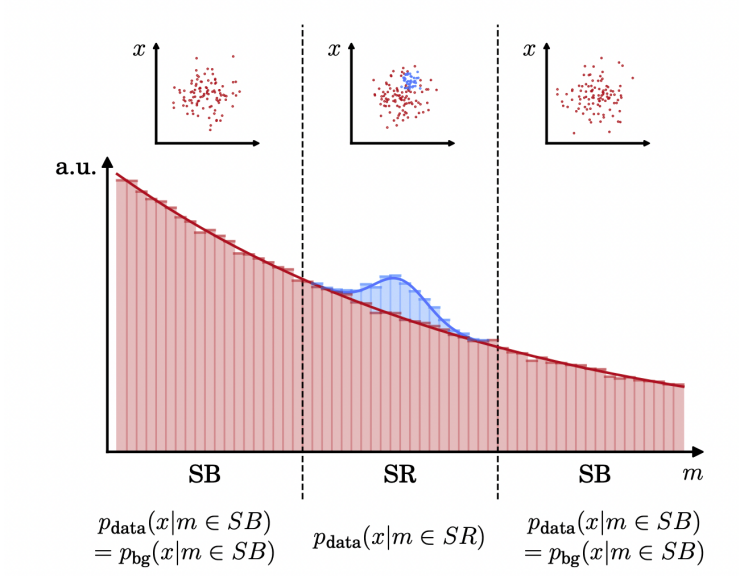}
    \caption{An illustration of the resonance-based overdensity scenario. The signal is localized in particular region of the resonant variable. Within that region there is an overdensity in the feature space from the signal events that is not present in the sidebands. Figure taken from \cite{Hallin:2021wme}.}
    \label{fig:resonant_overdensity}
\end{figure}

The assumptions of a narrow resonance search, in particular the localization of the signal, can be used to define the mixed samples needed for weak supervision.
Though typically one does not know the mass of the sought-after signal, 
one can guess a window in the mass distribution where one hopes that the signal is localized to. 
If a signal is present in the window, the sample of events inside the window will contain some non-zero signal fraction, while events outside of the window will not. 
Events in this window can then serve as the potentially signal-rich mixed sample. 
Events outside this region will have very similar background events to those inside the window, but may contain some differences. 
Assuming the features of the background change smoothly as a function of the resonant variable, the background in the signal region can be estimated via an interpolation from the sideband regions.
This is illustrated in Figure \ref{fig:resonant_overdensity}

Different approaches have been taken to use the sideband events to construct the background-rich mixed sample.
The first methods \cite{Collins:2018epr,Collins:2019jip, Amram:2020ykb} used a weighted sample of the events from the sidebands adjacent to the signal-window.
Others have improved upon this approach by training some sort of generative-like model from the sidebands and then interpolating it into the signal region. 
Samples are then drawn from this generative model to construct the background-rich sample. 
The first of these methods was CATHODE \cite{Hallin:2021wme, Hallin:2022eoq} which used a normalizing flow trained in the sidebands.
Other methods include SALAD \cite{Andreassen:2020nkr, Benkendorfer:2020gek} which uses simulation to help with the interpolation, CURTAINS \cite{Raine:2022hht,Sengupta:2023xqy} which 'transports' events from the sidebands, , and FETA \cite{Golling:2022nkl} which uses transport and simulation.
These methods seem to perform roughly similarly \cite{Golling:2023yjq}.
Other methods use density estimates to construct the likelihood ratio rather than a classifier \cite{ANODE,Stein:2020rou, RANODE}. 

The most well-studied case are dijet resonances, in which one looks for resonance decaying to two jets with anomalous substructure. 
There is a large phenomenology of jets which can produce substructure distinct from that of typical QCD jets. 
Jet substructure features have minimal correlation with the resonance mass, allowing one to perform a bump-hunt after an anomaly detection selection. 

There have been several results anomaly detection methods to dijet final states at the LHC \cite{ATLAS_cwola,ATLAS_Higgs_anomaly, CMS_CASE, CMS_CASE_MLG, CMS_Higgs_anomaly, ATLAS_semivisible_anomaly}.
Two recent searches from the CMS \cite{CMS_CASE, CMS_CASE_MLG} and ATLAS Collaborations \cite{ATLAS_anomaly} offer useful case studies for the application of these methods.
The CMS search deployed several different complementary anomaly detection strategies. An outlier detection method, based on a variational autoencoder (VAE) was employed. Three different weakly supervised strategies were used, each differing their construction of the two mixed samples: CWoLa Hunting \cite{Collins:2018epr,Collins:2019jip}, Tag N' Train \cite{Amram:2020ykb}, and CATHODE \cite{Hallin:2021wme}. \footnote{A semi-supervised method, QUAK~\cite{Park:2020pak}, was also employed. However as this is only a partially model-agnostic approach it will not be discussed further.}
The ATLAS search also deployed two different weakly supervised methods: SALAD \cite{Andreassen:2020nkr, Benkendorfer:2020gek} and CURTAINS \cite{Raine:2022hht}.
This variety of methods employed makes them useful case studies for general validation procedures for anomaly detection searches.

\subsection{Mass decorrelation}

In a bump-hunt analysis, spurious excesses can arise due to correlation between the anomaly score and the resonance mass if care is not taken.
This is a particular danger for weakly supervised methods which use windows in the resonance mass to define the `signal-like' sample for training.
If the features used for classification are correlated with the resonance mass, the learned anomaly score can become significantly correlated with the resonance mass as well.
Consequently, the selection criterion on the anomaly score can significant distort the resonance mass distribution.
In the extreme case the learned anomaly score can preferentially select events in the pre-defined mass window as signal-like, which would create a localized excess of background events mimicking a signal. 

Outlier detection methods can also distort the background mass distribution through a different mechanism. 
Since the outlier detection methods are defined to select rare events as anomalous, and invariant masses typically have steeply falling distributions, outlier detection methods will often preferentially select events in the high mass tails as more anomalous. 
This can significant distort the shape of the background mass distribution, making extraction of a signal difficult. 

Preventing these sorts of mass sculpting effects are therefore of the utmost importance in building a reliable resonant anomaly detection analysis. 
For the weakly supervised methods this can be achieved by using features uncorrelated with the resonance mass, and/or by improving the construction of the background-rich sample such that it has minimal kinematical differences with respect to the signal-rich sample. 
For outlier detection methods, the use of mass-uncorrelated features can also be employed.
Applying event weights to the training, to upweight events in the high mass tails to equalize their density respect to the copious low mass events can also be employed.
However in the case of steeply falling distributions whose density changes by orders of magnitude, the necessary large weights can become impractical for training. 
For outlier detection, because the learned anomaly score function is fixed, a post-training decorrelation between the anomaly score and the resonance mass can also be performed.
Such a strategy is much more difficult to deploy for weakly supervised methods, which have variable behavior.

The recent CMS dijet anomaly search \cite{CMS_CASE} employed several strategies to mitigate the issue of mass sculpting. 
The weakly supervised algorithms used a set of features only weakly correlated to the resonance mass, through a correlation with jet $p_T$. 
Their implementation of the CWoLa Hunting \cite{Collins:2018epr,Collins:2019jip} and TNT \cite{Amram:2020ykb} weakly supervised algorithms then used a jet $p_T$-based reweighting between the signal-rich and background-rich samples to eliminate any residual kinematic bias during the training.
The accuracy of the background-rich sample constructed by the CATHODE \cite{Hallin:2021wme} method was sufficient to not require this step. 
The VAE, which used inputs correlated with the jet $p_T$, applied a post-hoc correction to decorrelate the anomaly score from the resonance mass. 

\subsection{Validation Methods}

There are two key properties that an anomaly-detection-based analysis must satisfy to ensure a sound result.
The first is that the false-positive rate is controlled. 
That is, the employed strategy cannot be biased so as to report excesses at higher rates than expected under the background-only (null) hypothesis. 
The second is that the method is effective at finding anomalies: for some example signals of a realistic strength the method will result in a statistically significant rejection of the null hypothesis, ideally at the level of discovery ($5\sigma$). 
These are the same validations that must be done for any search strategy, however the validation of these properties are complicated in several ways by the anomaly detection methodology. 

It is worth distinguishing here between the validation of outlier detection methods and weakly supervised methods, the latter of which has unique challenges complicating its validation. 
For unsupervised methods, the anomaly detection algorithm is trained `ahead of time', without any dependence on the signal region data. The anomaly detection algorithm therefore `frozen', prior to unblinding, similarly to a supervised classifier.
For the weakly supervised methods, the training uses the signal region data, and the behavior of the anomaly detector will depend on the properties of the signal region data. 
This means the exact classification algorithm will not be fully known prior to unblinding. This means the entire training and signal extraction procedure must be validated in tandem, which is more involved than validating a frozen unsupervised model. 
This property also complicates the extraction of exclusion limits from weakly supervised searches, as discussed in Appendix A of Ref. \cite{CMS_CASE}.

\subsubsection{Validation of the Null Hypothesis}
Validating that an AD method is properly calibrated under the null hypothesis -- i.e., that it does not produce spurious excesses on background-only samples -- is crucially important
This validation can be done in simulation, in a data control region, or by using artificial data samples obtained from a generative model. 
Each strategy has a complementary set of benefits and limitations.
Therefore, the use of multiple strategies is desirable to ensure a robust validation of the method. 

\paragraph{Validation in simulation} The most straightforward validation strategy is the deployment of the algorithms on simulated Monte Carlo (MC) samples. 
These Monte Carlo samples should be as realistic to the application on data as possible.  
For the validation of weakly supervised methods, the common practice of using event weights to account for physics processes with different cross sections cannot be employed.
This is because these weights will affect the training dynamics and can influence the performance of the weakly supervised algorithm. 
For example, 10,000 events each with weight 0.01 will have much less statistical noise than 100 events with weight 1; the former may therefore be easier to learn from. 
When applied to real data all events will have weight 1, so the MC sample used for validation should as well.
To achieve this, events from different physics processes should be randomly sampled in proportion to their cross section. 
This ensures the right composition of events in the sample while maintaining the statistical properties that will be expected on data. 
The anomaly detection algorithms can then be run on this MC dataset without any signals to verify that the aforementioned mass sculpting does not occur.

While validation on the MC sample is very useful, it can also be limited in several respects. 
First, MC simulation is known to have mismodelings as compared to the data, particularly for QCD processes and jet substructure observables.
While these MC sets are being used only for validation purposes, and it is not required that they exactly match the data, features present in the data not captured by the MC, such as rare detector reconstruction effects or subtle kinematic correlations, could cause issues for the anomaly detection algorithms and would not show up in MC validations.
Additionally, algorithms which make explicit use of MC samples as part of their background estimate (such as SALAD \cite{Andreassen:2020nkr,Benkendorfer:2020gek}) would not be properly tested if the same MC sample is used for validation, as realistic data-MC differences would be absent. 
Systematically varied MC samples, perhaps originating from a different generator, could be used in these cases to approximate these affects, but it is likely this would not fully capture fully realistic data-MC differences. 
MC samples are also of a limited size, and for QCD processes with large cross sections one often has fewer simulated events available than will be present in the actual data sample. 
Running tests on these samples with limited size could fail to catch biases that are only apparent with larger statistics.
Though bootstrapping methods can be used, the limited sample size means it is difficult to run a large suite of toys to test the variability of the weakly supervised algorithms and ensure the null hypothesis is properly calibrated. 

\paragraph{Validation in data control regions}
For these reasons, additional validation strategies are desirable to further test these algorithms. 
An additional validation strategy is to run the algorithms on a data sample from a control region which is known from previous studies, or assumed based on physical arguments, to contain a negligible fraction of anomalies.
This control region should be designed to be as similar to the signal region as possible, both in terms of the number of events, and the distribution of kinematics and classification features, for a realistic test.
In a fully model independent search, it is difficult to define a control region which can be assumed to have negligible signal. 
Assumptions must be made about the characteristics of the signal to justifiably claim some portion of the data is signal depleted.
For the dijet resonance searches, such a control region was defined based on an kinematic property of each event: the rapidity separation between the two jets. 
Resonant signals targeted by the search would be produced via the $s$-channel and therefore lead to jets with a smaller rapidity separation than the dominant $t$-channel QCD background. 
This event property is orthogonal to the substructure features being used to look for anomalies and therefore retains the model-agnostic nature of the search.

It is possible that some other signal, not produced via the $s$-channel dijet topology, could indeed populate such a control region and be identified by the anomaly detection algorithm.
For example, a resonance decaying to three or more jets with anomalous substructure could result in two jets with high rapidity separation and thus populate the  control region in the dijet search. 
The anomaly detection methods would identify these anomalous events, leading to a rejection of the null hypothesis in the control region, complicating validation.
However, this is a general problem for all searches: one cannot rule out some other new physics model, differing from the one being searched for, contaminating a control region. 
Though we note because of the broad sensitivity of anomaly detection methods, they may be more sensitive to this possibility than other searches. 
Nevertheless, despite these philosophical issues, using such a control region still serves as very useful validations and was used in both the recent CMS and ATLAS searches.
In the case of any significant excess observed in such a control region, it would be necessary to study further whether it originates from a potentially genuine anomaly or a bias in the algorithm. 

Control regions also have practical limitations that motivate further validation strategies.
Kinematic differences between the control region and the signal region can affect the behavior of the algorithms, potentially masking potential bias. 
Additionally, control regions have finite numbers of events, limiting the ability to conduct to repeated toy experiments to test for statistical bias.
Additionally, for some searches, defining a suitable control region, with enough similarity to the signal region to serve as a useful validation, may be difficult or impossible. 

\paragraph{Validation on artificial samples}
The recent ATLAS dijet search \cite{ATLAS_anomaly} employed a novel strategy dubbed \texttt{DOWN-UP-SAMPLE} as a further form of validation. 
They trained a generative model on a random, small fraction of their signal region data.
This generative model is used to generate mock background samples which could be used used for pseudoexperiments to test for bias. 
The initial downsample of the signal region data effectively dilutes the presence of potential anomalies, which are assumed to be rare so as to not been found by previous searches.
This means the generative model will learn the background distribution only rather than additionally learning features of potential hidden anomalies. 
Tests in simulation were performed to confirm this property.
Because it is trained on the signal region data, the features learned by the generative model should have very realistic correlations, and new samples can easily generated to run a large number of pseudoexperiments of a realistic size.
This method allowed the ATLAS search to identify a significant bias in their method at low resonance masses which was not fully apparent from the other validation strategies. 

One potential limitation of this validation strategy is that if the anomaly detection technique itself uses a generative model, such as in CATHODE \cite{Hallin:2021wme} or CURTAINS \cite{Raine:2022hht}, it may have an easier time modeling this artificial background than the true data.
All generative models have inductive biases which lead to imperfections, such as preferring smoothly varying features or underestimating tails.
A dataset coming from a generative model would already have these imperfections, and therefore the generative model trained in the pseudoexperiment would have an easier time fully modeling such an artificial dataset than it will when applied to the true dataset.
This effect could be mitigated by training the initial generative model with a significantly different architecture so as not to align its inductive biases with the model trained in the pseudoexperiment. 

\subsubsection{Validation on true signals}
The other necessary validation is ensuring that the anomaly detection algorithm can successfully identify a set of true anomalies.
This necessarily involves selecting a set of benchmark signal models to be used for testing purposes. 

This set of benchmarks should be chosen to cover a wide phenomenological range, while still matching the target search topology.
For the dijet search this means considering many different models which produce two jets, but differing in the substructure of those two jets. 
The CMS search considered benchmark models in producing jets with between two and six `prongs' of energies and varied jet masses.
Employing such a wide range of signatures validates that the anomaly detection algorithm indeed have sensitivity to a broad class of models. 
It may also expose a class of models which the search is not especially sensitive, which can inform future efforts.
It is not required or expected that the anomaly detection method is sensitive to \emph{all} possible models.
Every methodology will employ a set of assumptions and choose a set of input features which will leave it insensitive to some models. 

\paragraph{Validation in simulation} Events from these signal models can be injected into the aforementioned realistic MC sample, and the entire anomaly detection analysis pipeline, from the AD selection to extraction of significance, performed on this dataset.
The number of signal events injected should be varied to test performance for different signal strengths.
This is particularly important for weakly supervised methods whose performance changes significantly as a function of the signal strength.
The signal injections sizes should still be kept in a reasonable range, reflecting realistic signal strengths that could potentially live within the data. 
Unless the search is probing an entirely new frontier or final state, extremely large signals would likely have been seen by previous searches, making performance validation in such a regime unnecessary. 

\paragraph{Validation in data}A validation of the AD methods identifying a true anomaly in data is also desirable.
This can be achieved by `rediscovering' rare standard model processes using the chosen anomaly detection. 
Such a validation requires there to be a standard model process similar enough to the characteristics of the sought after signal such that it fits within the scope of the AD method. 
Rediscovery of top quarks \cite{Knapp:2020dde} and the Upsilon \cite{Gambhir:2025afb} using anomaly detection methods have been demonstrated using open data from CMS.

In the case of the dijet search, there is no comparable standard model resonance decaying into two jets with anomalous substructure.
Instead, the recent CMS search instead used the pair production of high-momenta (boosted) top quarks as a validation process \cite{CMS_CASE_MLG}. 
This standard process has no central resonance, but does produce two jets with anomalous substructure.
Due to the lack of a central resonance, this validation required a modification of the training strategy of the weakly supervised methods, and a change to the final signal extraction procedure.
With these changes, the AD method was shown to be able to successfully enhance the fraction of boosted top quarks from a small component of the data sample ($< 1\%$) to a large, dominant portion of the selected 'anomalous' events. 
This procedure validates that the core basis behind the AD method functions as expected, which is an important.
But due to the changes required to probe a different signal topology, some details of the procedure, such as the signal extraction method, are not validated in this approach.
Therefore it should be viewed as complementary to the aforementioned validation in simulation. 
Any such validation strategy in data will be highly analysis dependent, and for many cases there may be no suitably similar standard model process to use. 

\subsection{Assessing performance}
How to assess the performance of the anomaly detection algorithms on the benchmark signals is also worth discussing.
Training a supervised classifier for each chosen signal can give a useful upper bound on the performance of the AD method. 
However, in most cases it is expected the AD method will fall well short of this.
The minimal baseline that the AD performance must clear is that it improves the sensitivity beyond an inclusive search method, which makes no anomaly-like selection.
If possible, the inclusive search should in all other ways be identical to the anomaly detection analysis; using the same inputs objects and statistical analysis procedure, just without the selection on an anomaly score. 
The significance improvement for the AD selection is approximately given by the signal efficiency of the selection ($\epsilon_s$) divided by the square root of the background efficiency ($\epsilon_b$), $\frac{\epsilon_s}{\sqrt{\epsilon_b}}$. 
This is a non-trivial hurdle to clear; sometimes even classifiers with seemingly moderate classification performance (AUC ${\sim}0.7$) do not result in significance improvements larger than one. 
Ideally, to demonstrate discovery potential, the anomaly detection algorithm should enhance significances which were below the level of evidence ($< 3 \sigma$) in the inclusive search to discovery-level ($5 \sigma$). 
The performance of AD algorithm can also be compared to other simple selection criteria appropriate to the search. 
For the dijet search the performance of the AD methods was compared to simple jet substructure selections commonly employed in searches.
For analyses focused on event-level anomalies, comparisons could be made to selections on global event quantities often employed in searches such as the scalar sum of object momenta (ST), the missing transverse energy (MET), or number of reconstructed objects in the event. 
Reporting the performance of these AD algorithms with respect to these baselines is important to clearly demonstrate the gains of the AD methods. 

The recent CMS search compared the performance of its AD methods to an inclusive search, as well as two sets of simple cut-based approaches, one targeting two-prong and the other targeting three-prong jets. 
Their performance was also compared to supervised classifiers trained with the same features as the AD methods.
These comparisons were done for all ${\sim}20$ benchmark signal models considered. 
Significance improvements with respect to the inclusive search as large as a factor of 6 were found for some signal models.
AD performance was generally a factor of two or more inferior to a supervised classifiers.
For several more challenging signals, no AD algorithm achieved improvement with respect to the inclusive search, indicating the need 
for further method development or new strategies. 

While the usage of benchmark signal models has great utility, they will likely not capture the full model space that an AD search will be sensitive to. 
If the signal models are available to the analyzers during development of the search methodology, one may worry that the AD algorithms have been tuned to give good performance on these models. 
If a significant tuning has occurred, the performance of the AD methods on unknown untested signals may be worse than the tested benchmarks. 
Therefore, in future searches it would be desirable to test the performance of AD algorithms on a set of `validation' signals which were not used during development of the AD algorithm and only tested at the time the algorithm is finalized and being applied to data. 
This would give external readers confidence on that the benchmark performance can indeed be extrapolated as rough estimates of sensitivity to untested signals. 
Ideally the nature of the signal would be blinded to the analyzers, but chosen by an external party to fall within the scope of the search.
This would be similar to the hidden signals used in the evaluation of prior community anomaly detection challenges~\cite{LHCO,darkmachines}.
However such a blinding may be difficult to achieve within experimental collaborations. 

\section{Search Interpretation and Follow Up}
\label{sec:interp}

A frequent question is how the results from such model-agnostic searches should be interpreted. 
As compared to conventional searches, these model-agnostic strategies have important differences in interpretation strategies, both in the case of a significant excess and in the reporting of exclusion limits.

In the case these model-agnostic methods results in a significant excess, steps must be taken to verify its nature.
Identifying that the features of the anomalous event match those of a realistic new particle, rather than features arising from subtle instrumental or reconstruction related effect, would help verify its physical origin.
The identification of the excess's phenomenological properties would also allow followup supervised analyses, targeting the specific signature matching the excess, to be employed to confirm the excess.

\subsection{Considerations for a follow-up targeted search}

It should be noted that a targeted search designed based on the results of an anomaly detection search on the same dataset, cannot have a well-calibrated p-value.
This is because the targeted search would not have been performed if there had not been an excess seen by the anomaly detection search, so it has an unquantifiable LEE.
However, if the targeted search is performed on an independent dataset, then it can have a well-calibrated p-value.
Validation from a targeted search performed on the same dataset can still provide value in the form of additional robustness checks and more easily interpretable methodology, but its p-value should not be reported.

Therefore, to enable a full verification, it may be beneficial for any anomaly detection strategy to pre-define a holdout dataset. 
This holdout dataset would be used only for analysis by the targeted search, in case of a significant excess seen by the anomaly detection method.
In ongoing experiments, such a holdout dataset may not be needed as any anomaly could be verified in future datasets, but for analyses with fixed datasets it is an important consideration.

The size of the holdout dataset is not an obvious choice. 
An optimized targeted analysis will have better sensitivity than a model-agnostic one, so reserving 50\% of the data for holdout would ensure the targeted follow up to have equivalent or greater statistical power than the model-agnostic one.
However, such a large holdout set would significantly reduce the statistical power of the initial model-agnostic strategy.
The ratio of sensitivities between a model-agnostic and targeted search strategy will depend on the nature of the anomaly and therefore cannot be determined a priori. 
If the targeted search had two times or more the sensitivity as a model-agnostic strategy, 
which is a reasonable assumption given current performance, a holdout size of 20\% would ensure enough statistical power for confirmation.
A holdout size in the range of 20\%-50\% therefore seems reasonable.

\subsection{Excess interpretation}

Before a targeted search can be performed, the nature of any observed anomaly must first be understood.
Several interpretation strategies have been investigated for both the two-sample test  and for the model-agnostic signal selection methods.

\paragraph{Feature distribution comparisons}
One of the simplest interpretation strategies is to compare the feature distributions of the highest anomaly score events from the signal excess were compared to those of typical background events.
This comparison allows a determination of what is unique about those events causing them to be identified as anomalous.
This is among the simplest interpretability strategies, but was shown to be effective in the CMS dijet search \cite{CMS_CASE}.
The method could be applied to two-sample tests as well, by plotting the distribution of features for data events with the highest likelihood ratio between data and reference.
The method relies on the analyzer selecting the features a priori to check the difference between the anomalous and regular events.
If the anomalous behavior involves the correlated behavior of two or more features, it may not be clearly apparent in simple one-dimensional distributions.
Similar checks could then be performed in two-dimensional distributions or on derived features which are functions of two or more of the input features.
Some trial and error may be required to find the features which properly characterize the anomaly.

\paragraph{Permutation feature importance}
For classifier-based approaches, including NPLM and weakly supervised approaches, it can be useful to interpret the behavior of the classifier itself to assess what it has learned.
In \cite{CMS_CASE}, a classifier-interpretation method based on the permutation feature importance \cite{permutation_score} was employed.
In the set of most anomalous events, a single chosen feature was permuted with those from a set of random other events, and the change in the classification score was computed.
The average change in the classification score across the variations was used to compute a feature importance score. 
Repeating this procedure for all the input features, and then comparing their scores, allows an assessment of which features were most important in the classification of the event as anomalous.

\paragraph{Active subspace method} In \cite{chakravarti2023model}, the authors propose to interpret a trained classifier by analyzing the information encoded in the gradient of a classifier’s decision function. 
When a classifier is trained for a model-agnostic two-sample test, its output defines a surface that separates the measured data from the background sample. 
The gradient of this surface with respect to the input features indicates the local directions in feature space that most influence the classifier’s decision, and thus the directions along which the two samples differ. 
By computing the covariance of these gradients over the data and performing an eigenvalue decomposition, the method extracts a set of \emph{active subspaces}: dominant linear combinations of features to which the classifier is most sensitive. 
Projecting events onto these subspaces returns low-dimensional, physically interpretable summaries that highlight which characteristics of the anomalous events drive the observed discrepancy. 
Such information can help determine whether the excess exhibits features consistent with a realistic new-physics signal or if it is more compatible with detector or reconstruction artifacts. 
The resulting interpretable directions can also guide the construction of targeted supervised analyses that specifically test the phenomenology suggested by the discrepancy that has been detected.

\paragraph{Reweighting the background sample} Another approach to understanding the nature of a discrepancy found by a classifier-based two-sample test was presented in \cite{DAgnolo:2018cun} in the context of the NPLM test. 
As discussed in Section~\ref{sec:nplm}, the classifier effectively learns a reweighting of the background distribution to match the distribution of the measured data (see Eq.~\eqref{eq:reweighting}). 
This learned reweighting function can be evaluated on the background events, either directly in terms of the input features or projected onto a high-level quantity such as an invariant mass. 
If the measured data and background sample are compatible, the fitted function $f_{\hat{w}}(z)$ will evaluate to approximately zero across the feature space. 
Conversely, deviations of $f_{\hat{w}}(z)$ from zero indicate regions where the data differ from the expected background, thereby providing a physically interpretable characterization of the discrepancy.

These strategies were found to be effective at determining the rough characteristics of the signal for both simulated signal injections and in the rediscovery of high momenta top quarks in the CMS dijet search \cite{CMS_CASE, CMS_CASE_MLG}.
Once the characteristics of the signal have been determined, a follow-up analysis which targets the specific signature of the excess can be performed. 
This targeted analysis will likely achieve higher sensitivity than the model-agnostic method, as the targeted analysis can employ optimized selection criteria, background estimation and signal extraction strategies.
If the properties excess  are not fully determined, several variations of the targeted analysis may be necessary to span the possibilities. 
As the targeted search was based on a the results of a search performed on the same dataset, it cannot be considered a fully blinded analysis and therefore faces an difficult-to-quantify LEE. As such, it should be used only to verify that a local excess indeed exists in the region reported by the model-agnostic strategy, but caution should be taken in reporting its global significance.
A full confirmation therefore requires the same targeted strategy to be performed on an orthogonal dataset. 

\subsection{Exclusion limits}
\label{subsec:limits}
Besides reporting any significant excesses, the other typical outcome of a new physics search are exclusion limits.
These exclusion limits rule out the existence of a particular new particle or phenomenon, under certain model assumptions, within a certain parameter space.
Such exclusion limits are used to the compare the sensitivities of different analyses, to assess the viability of different theories, and to identify gaps in coverage to be filled in by new analyses or experiments. 

For a model-agnostic search, reporting exclusion limits is complicated by two factors.
The first is that given the broad sensitivity of these analyses it is not possible to report exclusion limits which comprehensively summarize the sensitivity of the search. 
Even when a broad set of benchmark signal models are considered, there will undoubtedly be additional models to which the analysis is sensitive.
This issue could be ameliorated if such searches could easily be re-interpretable by the community, and the search's sensitivity to additional models could be tested after-the-fact by any interested party.
Unfortunately, since these analyses are based on non-standard methodologies, they are oftentimes much more difficult to re-interpret than standard ones. 

For the outlier-detection methods, reinterpretation seems more tractable. 
At a high-level, these analyses are functionally similar to standard search strategies, which have been reinterpreted by the community many times.
The only difference is that the selection criteria is based on unsupervised outlier detection models, whose efficiency is not easy to parameterize.
Instead of a universal parameterization, to estimate the selection efficiency on a new signal two alternate strategies are possible. 
The first is for the experiment to publicly release the machine learning model, so that it can be applied to additional signals by anyone in the community.
This strategy was adopted by a recent CMS search for resonances decaying to a Higgs plus an anomalous jet \cite{CMS_Higgs_anomaly_jet}.
One potential concern for this approach may be that reinterpretation efforts typically use external fast simulation packages such as Delphes \cite{Delphes}, which attempt to mimic realistic experimental simulations but are known to be imperfect. 
A machine learning model trained on real data or realistic full simulations may show quite different performance on a fast-simulation sample if there is significant mismodeling. 
This problem is not unique to anomaly detection, and methods such as surrogate models \cite{Bieringer:2024pzt} are being explored to accommodate these limitations. 
An alternative approach is to develop a dedicated portal maintained by the experiment to which new signals from the community can be submitted.
Submitted signals are then simulated by the experiment's full simulation pipeline, and then used to evaluate the efficiency.
This approach was adopted by a recent ATLAS unsupervised search \cite{ATLAS_two_body,ADFILTER}.

For weakly supervised methods and two-sample tests, reinterpretation is more challenging. 
These methods directly use information from the signal region for training. 
This means that the performance of the algorithm depends on the presence or absence of signal. 
If there is no signal presence in the dataset, the classifier will have poor performance at detecting a hypothetical signal. 
Thus directly reinterpreting a null-result search using the model which was trained on the data will yield very poor exclusion limits.
However, estimating what would have been observed by the analysis had there been a signal in the data, which is necessary to determine whether the signal is excluded by the observed results, requires an estimate of how the classifier would have performed in such a scenario. 
Therefore the classifier must be retrained for different signal strengths, to determine how its performance changes as a function of the amount of signal in the dataset.
This parameterized efficiency can then be used to extract exclusion limits. 
For weakly supervised methods, this procedure was first performed in the first ATLAS dijet search \cite{ATLAS_two_body}. 
A more comprehensive description of the procedure is described in detail in Appendix C of the recent CMS dijet search \cite{CMS_CASE}.

This type of procedure suffers from limitations. 
Firstly, it is quite computationally costly, as it requires running the full analysis procedure for many different signals and injection strengths and systematic variations.
As of now, this makes it impractical to do comprehensive scans of parameter space with this method.
The other major challenge is finding a realistic sample to use for these injection studies. 
For the weakly supervised searches, these injections have been done by injecting signal events directly into the data.
However, the presence of some unknown signal in the data, below the detection threshold of the analysis, could affect the performance of these injections. 
In Ref. \cite{CMS_CASE}, this was studied and determined not to produce undercoverage in the resulting exclusion limits to a significant degree, but this may not hold true in all cases. 
In the future the usage of proxy samples to perform these injections may be preferable, but it is difficult to construct samples that mimic the data to a sufficient degree and are known to be signal-free. 

For two-sample tests such as NPLM, the distribution of the test statistic under the null hypothesis is independent of the measured data if estimated with toy data samples (which are signal-free by construction). This means that the same null hypothesis can be tested against different sets of data with different amounts of signal injection to set exclusion limits, assuming that the background data are accurate. 

This procedure also has significant challenges with reinterpretation.
To evaluate the exclusion limit requires access to the full signal region data, which is often not released by experimental collaborations.
This means exclusion limits on additional signals cannot be readily derived by those outside experimental collaborations. 
One strategy to enable reinterpretation in this scenario would be to fully preserve the analysis workflow and use a portal-based approach to evaluate the analysis sensitivity to additional signals~\cite{FlexCAST}.
Developing faster approximate methods to estimate the analysis sensitivity would also be of great practical use.
Building upon methods from the statistics literature~\cite{janssen2008regions} which attempt to globally characterize what alternative hypothesis a particular test has power for could also be interesting for an assessment of more general exclusion limits.

\section{Conclusion and outlook}
\label{sec:conclusion}
These new model-agnostic search strategies have great promise to expand the discovery potential of modern experiments in fundamental physics. 
However, they have significant conceptual differences with respect to standard analyses, both in their goals, methods, validation strategies, and interpretation. 
We have endeavored to provide a concise review the main conceptual and practical features of these methods, with a particular focus on the strategies which have been employed to validate their effectiveness. 
At time of writing there have been only a handful of applications of these methods by experimental collaborations~\cite{ATLAS_cwola,ATLAS_two_body, ATLAS_Higgs_anomaly, ATLAS_anomaly, ATLAS_semivisible_anomaly, ATLAS_multilep_anomaly, CMS_CASE, CMS_CASE_MLG, CMS_Higgs_anomaly}.
We hope this document is a useful reference as the community familiarizes itself with these new methods and establishes best practices for their validation and interpretation. 

\section*{Acknowledgments}

This article is part of VERaIPHY (Validation \& Evaluation for Robust AI in PHYsics), a coordinated effort that unites researchers from fundamental physics, computer science and statistics to discuss principled frameworks for assessing the reliability and scientific validity of modern machine learning methods.

\paragraph{Author contributions}

OA and ML collaborated to draft Sections 1, 2, 5 and 6.
ML led the drafting of Section 3 and OA led Section 5.
MK provided guidance and advising. 

\paragraph{Funding information}
This report is available under Fermilab open technical publications as FERMILAB-PUB-26-0009-CMS-PPD. O.A. is supported by Fermi Forward Discovery Group, LLC under Contract No. 89243024CSC000002 with the U.S. Department of Energy, Office of Science, Office of High Energy Physics. M.L. acknowledges the financial support of the European Research Council (grant SLING 819789).




\bibliography{references.bib}

\begin{thebibliography}{100}
\providecommand{\url}[1]{\texttt{#1}}
\providecommand{\urlprefix}{URL }
\expandafter\ifx\csname urlstyle\endcsname\relax
  \providecommand{\doi}[1]{doi:\discretionary{}{}{}#1}\else
  \providecommand{\doi}{doi:\discretionary{}{}{}\begingroup
  \urlstyle{rm}\Url}\fi
\providecommand{\eprint}[2][]{\url{#2}}

\bibitem{Neyman:1933wgr}
J.~Neyman and E.~S. Pearson,
\newblock \emph{{On the Problem of the Most Efficient Tests of Statistical
  Hypotheses}},
\newblock Phil. Trans. Roy. Soc. Lond. A \textbf{231}(694-706), 289 (1933),
\newblock \doi{10.1098/rsta.1933.0009}.

\bibitem{Choudalakis:2011qn}
G.~Choudalakis,
\newblock \emph{{On hypothesis testing, trials factor, hypertests and the
  BumpHunter}},
\newblock In \emph{{PHYSTAT 2011}} (2011), \eprint{1101.0390}.

\bibitem{CMS:higgs}
V.~Khachatryan \emph{et~al.},
\newblock \emph{{Observation of the Diphoton Decay of the Higgs Boson and
  Measurement of Its Properties}},
\newblock Eur. Phys. J. C \textbf{74}(10), 3076 (2014),
\newblock \doi{10.1140/epjc/s10052-014-3076-z},
\newblock \eprint{1407.0558}.

\bibitem{ATLAS:higgs}
G.~Aad \emph{et~al.},
\newblock \emph{{Measurement of Higgs boson production in the diphoton decay
  channel in pp collisions at center-of-mass energies of 7 and 8 TeV with the
  ATLAS detector}},
\newblock Phys. Rev. D \textbf{90}(11), 112015 (2014),
\newblock \doi{10.1103/PhysRevD.90.112015},
\newblock \eprint{1408.7084}.

\bibitem{CMS:dijet}
A.~M. Sirunyan \emph{et~al.},
\newblock \emph{{Search for high mass dijet resonances with a new background
  prediction method in proton-proton collisions at $\sqrt{s} =$ 13 TeV}},
\newblock JHEP \textbf{05}, 033 (2020),
\newblock \doi{10.1007/JHEP05(2020)033},
\newblock \eprint{1911.03947}.

\bibitem{ATLAS:dijet}
G.~Aad \emph{et~al.},
\newblock \emph{{Search for new resonances in mass distributions of jet pairs
  using 139 fb$^{-1}$ of $pp$ collisions at $\sqrt{s}=13$ TeV with the ATLAS
  detector}},
\newblock JHEP \textbf{03}, 145 (2020),
\newblock \doi{10.1007/JHEP03(2020)145},
\newblock \eprint{1910.08447}.

\bibitem{D0:2000vuh}
B.~Abbott \emph{et~al.},
\newblock \emph{{Search for new physics in e\ensuremath{\mu}X data at D\O{}
  using SLEUTH: A quasi-model-independent search strategy for new physics}},
\newblock Phys. Rev. D \textbf{62}, 092004 (2000),
\newblock \doi{10.1103/PhysRevD.62.092004},
\newblock \eprint{hep-ex/0006011}.

\bibitem{H1:2004rlm}
A.~Aktas \emph{et~al.},
\newblock \emph{{A General search for new phenomena in ep scattering at HERA}},
\newblock Phys. Lett. B \textbf{602}, 14 (2004),
\newblock \doi{10.1016/j.physletb.2004.09.057},
\newblock \eprint{hep-ex/0408044}.

\bibitem{H1:2008aak}
F.~D. Aaron \emph{et~al.},
\newblock \emph{{A General Search for New Phenomena at HERA}},
\newblock Phys. Lett. B \textbf{674}, 257 (2009),
\newblock \doi{10.1016/j.physletb.2009.03.034},
\newblock \eprint{0901.0507}.

\bibitem{CDF:2007iou}
T.~Aaltonen \emph{et~al.},
\newblock \emph{{Model-Independent and Quasi-Model-Independent Search for New
  Physics at CDF}},
\newblock Phys. Rev. D \textbf{78}, 012002 (2008),
\newblock \doi{10.1103/PhysRevD.78.012002},
\newblock \eprint{0712.1311}.

\bibitem{CDF:2007ykt}
T.~Aaltonen \emph{et~al.},
\newblock \emph{{Model-Independent Global Search for New High-p(T) Physics at
  CDF}}  (2007),
\newblock \doi{10.2172/922303},
\newblock \eprint{0712.2534}.

\bibitem{CDF:2008voc}
T.~Aaltonen \emph{et~al.},
\newblock \emph{{Global Search for New Physics with 2.0 fb$^{-1}$ at CDF}},
\newblock Phys. Rev. D \textbf{79}, 011101 (2009),
\newblock \doi{10.1103/PhysRevD.79.011101},
\newblock \eprint{0809.3781}.

\bibitem{ATLAS:2018zdn}
M.~Aaboud \emph{et~al.},
\newblock \emph{{A strategy for a general search for new phenomena using
  data-derived signal regions and its application within the ATLAS
  experiment}},
\newblock Eur. Phys. J. C \textbf{79}(2), 120 (2019),
\newblock \doi{10.1140/epjc/s10052-019-6540-y},
\newblock \eprint{1807.07447}.

\bibitem{CMS:2020zjg}
A.~M. Sirunyan \emph{et~al.},
\newblock \emph{{MUSiC: a model-unspecific search for new physics in
  proton\textendash{}proton collisions at $\sqrt{s} = 13\,\text {TeV} $}},
\newblock Eur. Phys. J. C \textbf{81}(7), 629 (2021),
\newblock \doi{10.1140/epjc/s10052-021-09236-z},
\newblock \eprint{2010.02984}.

\bibitem{LHCO}
G.~Kasieczka \emph{et~al.},
\newblock \emph{{The LHC Olympics 2020 a community challenge for anomaly
  detection in high energy physics}},
\newblock Rept. Prog. Phys. \textbf{84}(12), 124201 (2021),
\newblock \doi{10.1088/1361-6633/ac36b9},
\newblock \eprint{2101.08320}.

\bibitem{darkmachines}
T.~Aarrestad \emph{et~al.},
\newblock \emph{{The Dark Machines Anomaly Score Challenge: Benchmark Data and
  Model Independent Event Classification for the Large Hadron Collider}},
\newblock SciPost Phys. \textbf{12}(1), 043 (2022),
\newblock \doi{10.21468/SciPostPhys.12.1.043},
\newblock \eprint{2105.14027}.

\bibitem{ATLAS_cwola}
G.~Aad \emph{et~al.},
\newblock \emph{{Dijet resonance search with weak supervision using
  $\sqrt{s}=13$ TeV $pp$ collisions in the ATLAS detector}},
\newblock Phys. Rev. Lett. \textbf{125}(13), 131801 (2020),
\newblock \doi{10.1103/PhysRevLett.125.131801},
\newblock \eprint{2005.02983}.

\bibitem{ATLAS_two_body}
G.~Aad \emph{et~al.},
\newblock \emph{{Search for New Phenomena in Two-Body Invariant Mass
  Distributions Using Unsupervised Machine Learning for Anomaly Detection at
  s=13{\,}{\,}TeV with the ATLAS Detector}},
\newblock Phys. Rev. Lett. \textbf{132}(8), 081801 (2024),
\newblock \doi{10.1103/PhysRevLett.132.081801},
\newblock \eprint{2307.01612}.

\bibitem{ATLAS_Higgs_anomaly}
G.~Aad \emph{et~al.},
\newblock \emph{{Anomaly detection search for new resonances decaying into a
  Higgs boson and a generic new particle $X$ in hadronic final states using
  $\sqrt{s} = 13$ TeV $pp$ collisions with the ATLAS detector}},
\newblock Phys. Rev. D \textbf{108}, 052009 (2023),
\newblock \doi{10.1103/PhysRevD.108.052009},
\newblock \eprint{2306.03637}.

\bibitem{ATLAS_anomaly}
G.~Aad \emph{et~al.},
\newblock \emph{{Weakly supervised anomaly detection for resonant new physics
  in the dijet final state using proton-proton collisions at s=13{\,}{\,}TeV
  with the ATLAS detector}},
\newblock Phys. Rev. D \textbf{112}(7), 072009 (2025),
\newblock \doi{10.1103/2yq5-vj59},
\newblock \eprint{2502.09770}.

\bibitem{ATLAS_semivisible_anomaly}
G.~Aad \emph{et~al.},
\newblock \emph{{Search for new physics in final states with semivisible jets
  or anomalous signatures using the ATLAS detector}},
\newblock Phys. Rev. D \textbf{112}(1), 012021 (2025),
\newblock \doi{10.1103/44zp-mh1q},
\newblock \eprint{2505.01634}.

\bibitem{ATLAS_multilep_anomaly}
G.~Aad \emph{et~al.},
\newblock \emph{{Search for Beyond the Standard Model physics with anomaly
  detection in multilepton final states in $pp$ collisions at $\sqrt{s}=13$ TeV
  with the ATLAS detector}}  (2025),
\newblock \eprint{2508.19778}.

\bibitem{CMS_CASE}
V.~Chekhovsky \emph{et~al.},
\newblock \emph{{Model-agnostic search for dijet resonances with anomalous jet
  substructure in proton{\textendash}proton collisions at $\sqrt{s}$ = 13
  TeV}},
\newblock Rept. Prog. Phys. \textbf{88}(6), 067802 (2025),
\newblock \doi{10.1088/1361-6633/add762},
\newblock \eprint{2412.03747}.

\bibitem{CMS_CASE_MLG}
A.~Hayrapetyan \emph{et~al.},
\newblock \emph{{Machine-learning techniques for model-independent searches in
  dijet final states}}  (2025),
\newblock \doi{10.5281/zenodo.16656501},
\newblock \eprint{2512.20395}.

\bibitem{CMS_Higgs_anomaly}
C.~Collaboration,
\newblock \emph{{Search for resonances decaying to a Higgs boson in the bb
  final state and an anomalous jet}},
\newblock CMS-PAS-B2G-24-015  (2025).

\bibitem{hastie2009elements}
T.~Hastie,
\newblock \emph{The elements of statistical learning: data mining, inference,
  and prediction} (2009).

\bibitem{Carzon:2025isu}
J.~Carzon, A.~Ghosh, R.~Izbicki, A.~Lee, L.~Masserano and D.~Whiteson,
\newblock \emph{{On Focusing Statistical Power for Searches and Measurements in
  Particle Physics}}  (2025),
\newblock \eprint{2507.17831}.

\bibitem{janssen2000global}
A.~Janssen,
\newblock \emph{Global power functions of goodness of fit tests},
\newblock Annals of Statistics pp. 239--253 (2000).

\bibitem{wasserman2013all}
L.~Wasserman,
\newblock \emph{All of statistics: a concise course in statistical inference},
\newblock Springer Science \& Business Media (2013).

\bibitem{friedman2003multivariate}
J.~Friedman,
\newblock \emph{On multivariate goodness-of-fit and two-sample testing},
\newblock Statistical Problems in Particle Physics, Astrophysics, and Cosmology
  p. 311 (2003).

\bibitem{lopez-paz2017revisiting}
D.~Lopez-Paz and M.~Oquab,
\newblock \emph{Revisiting classifier two-sample tests},
\newblock In \emph{International Conference on Learning Representations}
  (2017).

\bibitem{DAgnolo:2018cun}
R.~T. D'Agnolo and A.~Wulzer,
\newblock \emph{{Learning New Physics from a Machine}},
\newblock Phys. Rev. D \textbf{99}(1), 015014 (2019),
\newblock \doi{10.1103/PhysRevD.99.015014},
\newblock \eprint{1806.02350}.

\bibitem{DAgnolo:2019vbw}
R.~T. D'Agnolo, G.~Grosso, M.~Pierini, A.~Wulzer and M.~Zanetti,
\newblock \emph{{Learning multivariate new physics}},
\newblock Eur. Phys. J. C \textbf{81}(1), 89 (2021),
\newblock \doi{10.1140/epjc/s10052-021-08853-y},
\newblock \eprint{1912.12155}.

\bibitem{kim2021classification}
I.~Kim, A.~Ramdas, A.~Singh and L.~Wasserman,
\newblock \emph{Classification accuracy as a proxy for two-sample testing},
\newblock The Annals of Statistics \textbf{49}(1), 411 (2021).

\bibitem{Letizia:2022xbe}
M.~Letizia, G.~Losapio, M.~Rando, G.~Grosso, A.~Wulzer, M.~Pierini, M.~Zanetti
  and L.~Rosasco,
\newblock \emph{{Learning new physics efficiently with nonparametric methods}},
\newblock Eur. Phys. J. C \textbf{82}(10), 879 (2022),
\newblock \doi{10.1140/epjc/s10052-022-10830-y},
\newblock \eprint{2204.02317}.

\bibitem{chakravarti2023model}
P.~Chakravarti, M.~Kuusela, J.~Lei and L.~Wasserman,
\newblock \emph{Model-independent detection of new physics signals using
  interpretable semisupervised classifier tests},
\newblock The Annals of Applied Statistics \textbf{17}(4), 2759 (2023).

\bibitem{Grosso:2023scl}
G.~Grosso, M.~Letizia, M.~Pierini and A.~Wulzer,
\newblock \emph{{Goodness of fit by Neyman-Pearson testing}},
\newblock SciPost Phys. \textbf{16}, 123 (2024),
\newblock \doi{10.21468/SciPostPhys.16.5.123},
\newblock \eprint{2305.14137}.

\bibitem{Grossi:2025pmm}
S.~Grossi, M.~Letizia and R.~Torre,
\newblock \emph{{Comparing Generative Models with the New Physics Learning
  Machine}}  (2025),
\newblock \eprint{2508.02275}.

\bibitem{JMLR:v13:gretton12a}
A.~Gretton, K.~M. Borgwardt, M.~J. Rasch, B.~Sch{{\"o}}lkopf and A.~Smola,
\newblock \emph{A kernel two-sample test},
\newblock Journal of Machine Learning Research \textbf{13}(25), 723 (2012).

\bibitem{chatalic2025efficient}
A.~Chatalic, M.~Letizia, N.~Schreuder and L.~Rosasco,
\newblock \emph{An efficient permutation-based kernel two-sample test},
\newblock arXiv preprint arXiv:2502.13570  (2025).

\bibitem{ramdas2017wasserstein}
A.~Ramdas, N.~Garc{\'\i}a~Trillos and M.~Cuturi,
\newblock \emph{On wasserstein two-sample testing and related families of
  nonparametric tests},
\newblock Entropy \textbf{19}(2), 47 (2017).

\bibitem{Grossi:2024axb}
S.~Grossi, M.~Letizia and R.~Torre,
\newblock \emph{{Refereeing the referees: evaluating two-sample tests for
  validating generators in precision sciences}},
\newblock Mach. Learn. Sci. Tech. \textbf{6}(1), 015052 (2025),
\newblock \doi{10.1088/2632-2153/adb3ee},
\newblock \eprint{2409.16336}.

\bibitem{tran2025minimax}
B.~T. Tran and N.~Schreuder,
\newblock \emph{Minimax-optimal two-sample test with sliced wasserstein},
\newblock arXiv preprint arXiv:2510.27498  (2025).

\bibitem{biggs2023mmd}
F.~Biggs, A.~Schrab and A.~Gretton,
\newblock \emph{Mmd-fuse: Learning and combining kernels for two-sample testing
  without data splitting},
\newblock Advances in Neural Information Processing Systems \textbf{36}, 75151
  (2023).

\bibitem{schrab2023mmd}
A.~Schrab, I.~Kim, M.~Albert, B.~Laurent, B.~Guedj and A.~Gretton,
\newblock \emph{Mmd aggregated two-sample test},
\newblock Journal of Machine Learning Research \textbf{24}(194), 1 (2023).

\bibitem{Grosso:2024wjt}
G.~Grosso and M.~Letizia,
\newblock \emph{{Multiple testing for signal-agnostic searches for new physics
  with machine learning}},
\newblock Eur. Phys. J. C \textbf{85}(1), 4 (2025),
\newblock \doi{10.1140/epjc/s10052-024-13722-5},
\newblock \eprint{2408.12296}.

\bibitem{Lester:2021aks}
C.~G. Lester and R.~Tombs,
\newblock \emph{{Using unsupervised learning to detect broken symmetries, with
  relevance to searches for parity violation in nature. (Previously: ''Stressed
  GANs snag desserts'')}}  (2021),
\newblock \eprint{2111.00616}.

\bibitem{Tombs:2021wae}
R.~Tombs and C.~G. Lester,
\newblock \emph{{A method to challenge symmetries in data with self-supervised
  learning}},
\newblock JINST \textbf{17}(08), P08024 (2022),
\newblock \doi{10.1088/1748-0221/17/08/P08024},
\newblock \eprint{2111.05442}.

\bibitem{Taylor:2023deh}
P.~L. Taylor, M.~Craigie and Y.-S. Ting,
\newblock \emph{{Unsupervised searches for cosmological parity violation: An
  investigation with convolutional neural networks}},
\newblock Phys. Rev. D \textbf{109}(8), 083518 (2024),
\newblock \doi{10.1103/PhysRevD.109.083518},
\newblock \eprint{2312.09287}.

\bibitem{Craigie:2024bhk}
M.~Craigie, P.~L. Taylor, Y.-S. Ting, C.~Cuesta-Lazaro, R.~Ruggeri and T.~M.
  Davis,
\newblock \emph{{Unsupervised Searches for Cosmological Parity Violation:
  Improving Detection Power with the Neural Field Scattering Transform}}
  (2024),
\newblock \eprint{2405.13083}.

\bibitem{Park:2020pak}
S.~E. Park, D.~Rankin, S.-M. Udrescu, M.~Yunus and P.~Harris,
\newblock \emph{{Quasi Anomalous Knowledge: Searching for new physics with
  embedded knowledge}},
\newblock JHEP \textbf{21}, 030 (2020),
\newblock \doi{10.1007/JHEP06(2021)030},
\newblock \eprint{2011.03550}.

\bibitem{Cheng:2024yig}
C.~L. Cheng, G.~Singh and B.~Nachman,
\newblock \emph{{Incorporating Physical Priors into Weakly Supervised Anomaly
  Detection}},
\newblock Phys. Rev. Lett. \textbf{135}(2), 021801 (2025),
\newblock \doi{10.1103/8259-wt5p},
\newblock \eprint{2405.08889}.

\bibitem{Heimel:2018mkt}
T.~Heimel, G.~Kasieczka, T.~Plehn and J.~M. Thompson,
\newblock \emph{{QCD or What?}},
\newblock SciPost Phys. \textbf{6}(3), 030 (2019),
\newblock \doi{10.21468/SciPostPhys.6.3.030},
\newblock \eprint{1808.08979}.

\bibitem{Farina:2018fyg}
M.~Farina, Y.~Nakai and D.~Shih,
\newblock \emph{{Searching for New Physics with Deep Autoencoders}},
\newblock Phys. Rev. D \textbf{101}(7), 075021 (2020),
\newblock \doi{10.1103/PhysRevD.101.075021},
\newblock \eprint{1808.08992}.

\bibitem{kingma2014autoencoding}
D.~P. Kingma and M.~Welling,
\newblock \emph{Auto-encoding variational {Bayes}},
\newblock In \emph{Proc. 2nd Int. Conf. on Learning Representations} (2014),
  \eprint{1312.6114}.

\bibitem{Cerri:2018anq}
O.~Cerri, T.~Q. Nguyen, M.~Pierini, M.~Spiropulu and J.-R. Vlimant,
\newblock \emph{{Variational Autoencoders for New Physics Mining at the Large
  Hadron Collider}},
\newblock JHEP \textbf{05}, 036 (2019),
\newblock \doi{10.1007/JHEP05(2019)036},
\newblock \eprint{1811.10276}.

\bibitem{rezende2016variational}
D.~J. Rezende and S.~Mohamed,
\newblock \emph{Variational inference with normalizing flows},
\newblock in Proc. 32nd Int. Conf. on Machine Learning - vol. 37  (2016),
\newblock \doi{10.5555/3045118.3045281},
\newblock \eprint{1505.05770}.

\bibitem{lipman2024flowmatchingguidecode}
Y.~Lipman, M.~Havasi, P.~Holderrieth, N.~Shaul, M.~Le, B.~Karrer, R.~T.~Q.
  Chen, D.~Lopez-Paz, H.~Ben-Hamu and I.~Gat,
\newblock \emph{Flow matching guide and code}  (2024),
\newblock \eprint{2412.06264}.

\bibitem{Mikuni:2023tok}
V.~Mikuni and B.~Nachman,
\newblock \emph{{High-dimensional and permutation invariant anomaly
  detection}},
\newblock SciPost Phys. \textbf{16}(3), 062 (2024),
\newblock \doi{10.21468/SciPostPhys.16.3.062},
\newblock \eprint{2306.03933}.

\bibitem{Vaselli:2025zkl}
F.~Vaselli, M.~Pierini, M.~M. Glowacki, T.~Aarrestad, K.~Govorkova, V.~Loncar,
  D.~Danopoulos and F.~Pantaleo,
\newblock \emph{{It's not a FAD: first results in using Flows for unsupervised
  Anomaly Detection at 40 MHz at the Large Hadron Collider}},
\newblock In \emph{{ML4Jets 2025}},
\newblock \doi{10.48550/arXiv.2508.11594} (2025), \eprint{2508.11594}.

\bibitem{Kasieczka:2022naq}
G.~Kasieczka, R.~Mastandrea, V.~Mikuni, B.~Nachman, M.~Pettee and D.~Shih,
\newblock \emph{{Anomaly Detection under Coordinate Transformations}},
\newblock Phys.Rev.D \textbf{107}, 015009 (2022),
\newblock \doi{10.1103/PhysRevD.107.015009},
\newblock \eprint{2209.06225}.

\bibitem{OOD_answer}
Y.~L. Li, D.~Lu, P.~Kirichenko, S.~Qiu, T.~G.~J. Rudner, C.~B. Bruss and A.~G.
  Wilson,
\newblock \emph{Out-of-distribution detection methods answer the wrong
  questions} (2025), \eprint{2507.01831}.

\bibitem{Gordon_OOD}
P.~Kirichenko, P.~Izmailov and A.~G. Wilson,
\newblock \emph{Why normalizing flows fail to detect out-of-distribution data},
\newblock NIPS '20. Curran Associates Inc.,
\newblock ISBN 9781713829546 (2020).

\bibitem{serra_OOD}
J.~Serrà, D.~Álvarez, V.~Gómez, O.~Slizovskaia, J.~F. Núñez and J.~Luque,
\newblock \emph{Input complexity and out-of-distribution detection with
  likelihood-based generative models}  (2020),
\newblock \eprint{1909.11480}.

\bibitem{Buss:2022lxw}
T.~Buss, B.~M. Dillon, T.~Finke, M.~Kr\"amer, A.~Morandini, A.~M\"uck,
  I.~Oleksiyuk and T.~Plehn,
\newblock \emph{{What's Anomalous in LHC Jets?}},
\newblock SciPost Phys. \textbf{15}, 168 (2022),
\newblock \doi{10.21468/SciPostPhys.15.4.168},
\newblock \eprint{2202.00686}.

\bibitem{Dillon:2022mkq}
B.~M. Dillon, L.~Favaro, T.~Plehn, P.~Sorrenson and M.~Kr\"amer,
\newblock \emph{{A Normalized Autoencoder for LHC Triggers}},
\newblock SciPost Phys.Core \textbf{6}, 074 (2022),
\newblock \doi{10.21468/SciPostPhysCore.6.4.074},
\newblock \eprint{2206.14225}.

\bibitem{CMS_WNAE}
A.~Hayrapetyan \emph{et~al.},
\newblock \emph{{Wasserstein normalized autoencoder for anomaly detection}}
  (2025),
\newblock \eprint{2510.02168}.

\bibitem{Metodiev:2017vrx}
E.~M. Metodiev, B.~Nachman and J.~Thaler,
\newblock \emph{{Classification without labels: Learning from mixed samples in
  high energy physics}},
\newblock JHEP \textbf{10}, 174 (2017),
\newblock \doi{10.1007/JHEP10(2017)174},
\newblock \eprint{1708.02949}.

\bibitem{Collins:2018epr}
J.~H. Collins, K.~Howe and B.~Nachman,
\newblock \emph{{Anomaly Detection for Resonant New Physics with Machine
  Learning}},
\newblock Phys. Rev. Lett. \textbf{121}(24), 241803 (2018),
\newblock \doi{10.1103/PhysRevLett.121.241803},
\newblock \eprint{1805.02664}.

\bibitem{Collins:2019jip}
J.~H. Collins, K.~Howe and B.~Nachman,
\newblock \emph{{Extending the search for new resonances with machine
  learning}},
\newblock Phys. Rev. \textbf{D99}(1), 014038 (2019),
\newblock \doi{10.1103/PhysRevD.99.014038},
\newblock \eprint{1902.02634}.

\bibitem{Amram:2020ykb}
O.~Amram and C.~M. Suarez,
\newblock \emph{{Tag N{\textquoteright} Train: a technique to train improved
  classifiers on unlabeled data}},
\newblock JHEP \textbf{01}, 153 (2021),
\newblock \doi{10.1007/JHEP01(2021)153},
\newblock \eprint{2002.12376}.

\bibitem{ANODE}
B.~Nachman and D.~Shih,
\newblock \emph{{Anomaly Detection with Density Estimation}},
\newblock Phys. Rev. D \textbf{101}, 075042 (2020),
\newblock \doi{10.1103/PhysRevD.101.075042},
\newblock \eprint{2001.04990}.

\bibitem{Hallin:2021wme}
A.~Hallin, J.~Isaacson, G.~Kasieczka, C.~Krause, B.~Nachman, T.~Quadfasel,
  M.~Schlaffer, D.~Shih and M.~Sommerhalder,
\newblock \emph{{Classifying Anomalies THrough Outer Density Estimation
  (CATHODE)}},
\newblock Phys.Rev.D \textbf{106}, 055006 (2021),
\newblock \doi{10.1103/PhysRevD.106.055006},
\newblock \eprint{2109.00546}.

\bibitem{Hallin:2022eoq}
A.~Hallin, G.~Kasieczka, T.~Quadfasel, D.~Shih and M.~Sommerhalder,
\newblock \emph{{Resonant anomaly detection without background sculpting}},
\newblock Phys.Rev.D \textbf{107}, 114012 (2022),
\newblock \doi{10.1103/PhysRevD.107.114012},
\newblock \eprint{2210.14924}.

\bibitem{Stein:2020rou}
G.~Stein, U.~Seljak and B.~Dai,
\newblock \emph{{Unsupervised in-distribution anomaly detection of new physics
  through conditional density estimation}},
\newblock In \emph{{34th Conference on Neural Information Processing Systems}}
  (2020), \eprint{2012.11638}.

\bibitem{Andreassen:2020nkr}
A.~Andreassen, B.~Nachman and D.~Shih,
\newblock \emph{{Simulation Assisted Likelihood-free Anomaly Detection}},
\newblock Phys. Rev. D \textbf{101}(9), 095004 (2020),
\newblock \doi{10.1103/PhysRevD.101.095004},
\newblock \eprint{2001.05001}.

\bibitem{Benkendorfer:2020gek}
K.~Benkendorfer, L.~L. Pottier and B.~Nachman,
\newblock \emph{{Simulation-assisted decorrelation for resonant anomaly
  detection}},
\newblock Phys. Rev. D \textbf{104}(3), 035003 (2021),
\newblock \doi{10.1103/PhysRevD.104.035003},
\newblock \eprint{2009.02205}.

\bibitem{Raine:2022hht}
J.~A. Raine, S.~Klein, D.~Sengupta and T.~Golling,
\newblock \emph{{CURTAINs for your Sliding Window: Constructing Unobserved
  Regions by Transforming Adjacent Intervals}},
\newblock Front.Big Data \textbf{6}, 899345 (2022),
\newblock \doi{10.3389/fdata.2023.899345},
\newblock \eprint{2203.09470}.

\bibitem{Golling:2022nkl}
T.~Golling, S.~Klein, R.~Mastandrea and B.~Nachman,
\newblock \emph{{Flow-enhanced transportation for anomaly detection}},
\newblock Phys. Rev. D \textbf{107}(9), 096025 (2023),
\newblock \doi{10.1103/PhysRevD.107.096025},
\newblock \eprint{2212.11285}.

\bibitem{Kamenik:2022qxs}
J.~F. Kamenik and M.~Szewc,
\newblock \emph{{Null hypothesis test for anomaly detection}},
\newblock Phys. Lett. B \textbf{840}, 137836 (2023),
\newblock \doi{10.1016/j.physletb.2023.137836},
\newblock \eprint{2210.02226}.

\bibitem{Chen:2022suv}
M.~F. Chen, B.~Nachman and F.~Sala,
\newblock \emph{{Resonant anomaly detection with multiple reference datasets}},
\newblock JHEP \textbf{07}, 188 (2023),
\newblock \doi{10.1007/JHEP07(2023)188},
\newblock \eprint{2212.10579}.

\bibitem{Sengupta:2023xqy}
D.~Sengupta, S.~Klein, J.~A. Raine and T.~Golling,
\newblock \emph{{CURTAINs flows for flows: Constructing unobserved regions with
  maximum likelihood estimation}},
\newblock SciPost Phys. \textbf{17}(2), 046 (2024),
\newblock \doi{10.21468/SciPostPhys.17.2.046},
\newblock \eprint{2305.04646}.

\bibitem{Finke:2022lsu}
T.~Finke, M.~Kr\"amer, M.~Lipp and A.~M\"uck,
\newblock \emph{{Boosting mono-jet searches with model-agnostic machine
  learning}},
\newblock JHEP \textbf{08}, 015 (2022),
\newblock \doi{10.1007/JHEP08(2022)015},
\newblock \eprint{2204.11889}.

\bibitem{Bickendorf:2023nej}
G.~Bickendorf, M.~Drees, G.~Kasieczka, C.~Krause and D.~Shih,
\newblock \emph{{Combining resonant and tail-based anomaly detection}},
\newblock Phys. Rev. D \textbf{109}(9), 096031 (2024),
\newblock \doi{10.1103/PhysRevD.109.096031},
\newblock \eprint{2309.12918}.

\bibitem{Golling:2023yjq}
T.~Golling, G.~Kasieczka, C.~Krause, R.~Mastandrea, B.~Nachman, J.~A. Raine,
  D.~Sengupta, D.~Shih and M.~Sommerhalder,
\newblock \emph{{The interplay of machine learning-based resonant anomaly
  detection methods}},
\newblock Eur. Phys. J. C \textbf{84}(3), 241 (2024),
\newblock \doi{10.1140/epjc/s10052-024-12607-x},
\newblock \eprint{2307.11157}.

\bibitem{Buhmann:2023acn}
E.~Buhmann, C.~Ewen, G.~Kasieczka, V.~Mikuni, B.~Nachman and D.~Shih,
\newblock \emph{{Full phase space resonant anomaly detection}},
\newblock Phys. Rev. D \textbf{109}(5), 055015 (2024),
\newblock \doi{10.1103/PhysRevD.109.055015},
\newblock \eprint{2310.06897}.

\bibitem{SIGMA}
R.~Das and D.~Shih,
\newblock \emph{{Single interpolated generative model for anomalies}},
\newblock Phys. Rev. D \textbf{112}(7), 074040 (2025),
\newblock \doi{10.1103/rj53-2x6j},
\newblock \eprint{2410.20537}.

\bibitem{RANODE}
R.~Das, G.~Kasieczka and D.~Shih,
\newblock \emph{{Residual ANODE}}  (2023),
\newblock \eprint{2312.11629}.

\bibitem{Bai:2023yyy}
K.~Bai, R.~Mastandrea and B.~Nachman,
\newblock \emph{{Non-resonant anomaly detection with background
  extrapolation}},
\newblock JHEP \textbf{04}, 059 (2024),
\newblock \doi{10.1007/JHEP04(2024)059},
\newblock \eprint{2311.12924}.

\bibitem{Kasieczka:2024lxf}
G.~Kasieczka, J.~A. Raine, D.~Shih and A.~Upadhyay,
\newblock \emph{{Complete Optimal Non-Resonant Anomaly Detection}}  (2024),
\newblock \eprint{2404.07258}.

\bibitem{Shih:2021kbt}
D.~Shih, M.~R. Buckley, L.~Necib and J.~Tamanas,
\newblock \emph{{Via Machinae: Searching for Stellar Streams using Unsupervised
  Machine Learning}},
\newblock Mon.Not.Roy.Astron.Soc. \textbf{509}, 5992 (2021),
\newblock \doi{10.1093/mnras/stab3372},
\newblock \eprint{2104.12789}.

\bibitem{shih2023machinae}
D.~Shih, M.~R. Buckley and L.~Necib,
\newblock \emph{Via machinae 2.0: Full-sky, model-agnostic search for stellar
  streams in gaia dr2} (2023), \eprint{2303.01529}.

\bibitem{pettee2023weaklysupervised}
M.~Pettee, S.~Thanvantri, B.~Nachman, D.~Shih, M.~R. Buckley and J.~H. Collins,
\newblock \emph{Weakly-supervised anomaly detection in the milky way} (2023),
  \eprint{2305.03761}.

\bibitem{sengupta2024skycurtains}
D.~Sengupta, S.~Mulligan, D.~Shih, J.~A. Raine and T.~Golling,
\newblock \emph{{skycurtains: model-agnostic search for stellar streams with
  Gaia data}},
\newblock Mon. Not. Roy. Astron. Soc. \textbf{536}(2), 1104 (2024),
\newblock \doi{10.1093/mnras/stae2570},
\newblock \eprint{2405.12131}.

\bibitem{WeakSup_LEE}
P.~Shyamsundar, N.~Smith and M.~Szewc,
\newblock \emph{Unaccounted-for look-elsewhere effect in k-fold cross adaptive
  anomaly searches} (2025).

\bibitem{look_everywhere}
M.~Hein, B.~Nachman and D.~Shih,
\newblock \emph{{Look everywhere effects in anomaly detection}}  (2025),
\newblock \eprint{2512.13787}.

\bibitem{Baker:1983tu}
S.~Baker and R.~D. Cousins,
\newblock \emph{{Clarification of the Use of Chi Square and Likelihood
  Functions in Fits to Histograms}},
\newblock Nucl. Instrum. Meth. \textbf{221}, 437 (1984),
\newblock \doi{10.1016/0167-5087(84)90016-4}.

\bibitem{dAgnolo:2021aun}
R.~T. d'Agnolo, G.~Grosso, M.~Pierini, A.~Wulzer and M.~Zanetti,
\newblock \emph{{Learning new physics from an imperfect machine}},
\newblock Eur. Phys. J. C \textbf{82}(3), 275 (2022),
\newblock \doi{10.1140/epjc/s10052-022-10226-y},
\newblock \eprint{2111.13633}.

\bibitem{ParticleDataGroup:2024cfk}
S.~Navas \emph{et~al.},
\newblock \emph{{Review of particle physics}},
\newblock Phys. Rev. D \textbf{110}(3), 030001 (2024),
\newblock \doi{10.1103/PhysRevD.110.030001}.

\bibitem{Belis:2023mqs}
V.~Belis, P.~Odagiu and T.~K. Aarrestad,
\newblock \emph{{Machine learning for anomaly detection in particle physics}},
\newblock Rev. Phys. \textbf{12}, 100091 (2024),
\newblock \doi{10.1016/j.revip.2024.100091},
\newblock \eprint{2312.14190}.

\bibitem{CMS:2024nsz}
V.~Chekhovsky \emph{et~al.},
\newblock \emph{{Model-agnostic search for dijet resonances with anomalous jet
  substructure in proton{\textendash}proton collisions at $\sqrt{s}$ = 13
  TeV}},
\newblock Rept. Prog. Phys. \textbf{88}(6), 067802 (2025),
\newblock \doi{10.1088/1361-6633/add762},
\newblock \eprint{2412.03747}.

\bibitem{ATLAS:2020iwa}
G.~Aad \emph{et~al.},
\newblock \emph{{Dijet resonance search with weak supervision using
  $\sqrt{s}=13$ TeV $pp$ collisions in the ATLAS detector}},
\newblock Phys. Rev. Lett. \textbf{125}(13), 131801 (2020),
\newblock \doi{10.1103/PhysRevLett.125.131801},
\newblock \eprint{2005.02983}.

\bibitem{Metzger:2025ecl}
K.~Metzger, L.~Xu, M.~Sodini, T.~K. Arrestad, K.~Govorkova, G.~Grosso and
  P.~Harris,
\newblock \emph{{Anomaly-preserving contrastive neural embeddings for
  end-to-end model-independent searches at the LHC}},
\newblock Phys. Rev. D \textbf{112}(7), 072011 (2025),
\newblock \doi{10.1103/5n77-ynsp},
\newblock \eprint{2502.15926}.

\bibitem{Grosso:2023ltd}
G.~Grosso, N.~Lai, M.~Letizia, J.~Pazzini, M.~Rando, L.~Rosasco, A.~Wulzer and
  M.~Zanetti,
\newblock \emph{{Fast kernel methods for data quality monitoring as a
  goodness-of-fit test}},
\newblock Mach. Learn. Sci. Tech. \textbf{4}(3), 035029 (2023),
\newblock \doi{10.1088/2632-2153/acebb7},
\newblock \eprint{2303.05413}.

\bibitem{Cappelli:2025myc}
P.~Cappelli, G.~Grosso, M.~Letizia, H.~Reyes-Gonz{\'a}lez and M.~Zanetti,
\newblock \emph{{Learning to Validate Generative Models: a Goodness-of-Fit
  Approach}}  (2025),
\newblock \eprint{2511.09118}.

\bibitem{Knapp:2020dde}
O.~Knapp, O.~Cerri, G.~Dissertori, T.~Q. Nguyen, M.~Pierini and J.-R. Vlimant,
\newblock \emph{{Adversarially Learned Anomaly Detection on CMS Open Data:
  re-discovering the top quark}},
\newblock Eur. Phys. J. Plus \textbf{136}(2), 236 (2021),
\newblock \doi{10.1140/epjp/s13360-021-01109-4},
\newblock \eprint{2005.01598}.

\bibitem{Gambhir:2025afb}
R.~Gambhir, R.~Mastandrea, B.~Nachman and J.~Thaler,
\newblock \emph{{Isolating Unisolated Upsilons with Anomaly Detection in CMS
  Open Data}},
\newblock Phys. Rev. Lett. \textbf{135}(2), 021902 (2025),
\newblock \doi{10.1103/vvv3-5kkl},
\newblock \eprint{2502.14036}.

\bibitem{permutation_score}
A.~Altmann, L.~Toloşi, O.~Sander and T.~Lengauer,
\newblock \emph{Permutation importance: a corrected feature importance
  measure},
\newblock Bioinformatics \textbf{26}(10), 1340 (2010),
\newblock \doi{10.1093/bioinformatics/btq134}.

\bibitem{CMS_Higgs_anomaly_jet}
A.~Hayrapetyan \emph{et~al.},
\newblock \emph{{Search for resonances decaying to an anomalous jet and a Higgs
  boson in proton-proton collisions at $\sqrt{s}$ = 13 TeV}}  (2025),
\newblock \eprint{2509.13635}.

\bibitem{Delphes}
J.~de~Favereau, C.~Delaere, P.~Demin, A.~Giammanco, V.~Lema{\^\i}tre,
  A.~Mertens and M.~Selvaggi,
\newblock \emph{{DELPHES 3, A modular framework for fast simulation of a
  generic collider experiment}},
\newblock JHEP \textbf{02}, 057 (2014),
\newblock \doi{10.1007/JHEP02(2014)057},
\newblock \eprint{1307.6346}.

\bibitem{Bieringer:2024pzt}
S.~Bieringer, G.~Kasieczka, J.~Kieseler and M.~Trabs,
\newblock \emph{{Classifier surrogates: sharing AI-based searches with the
  world}},
\newblock Eur. Phys. J. C \textbf{84}(9), 972 (2024),
\newblock \doi{10.1140/epjc/s10052-024-13353-w},
\newblock \eprint{2402.15558}.

\bibitem{ADFILTER}
S.~V. Chekanov, W.~Islam, R.~Zhang and N.~Luongo,
\newblock \emph{{ADFilter{\textemdash}A Web Tool for New Physics Searches with
  Autoencoder-Based Anomaly Detection Using Deep Unsupervised Neural
  Networks}},
\newblock Information \textbf{16}(4), 258 (2025),
\newblock \doi{10.3390/info16040258},
\newblock \eprint{2409.03065}.

\bibitem{FlexCAST}
B.~Nachman and D.~Noll,
\newblock \emph{{FlexCAST: Enabling Flexible Scientific Data Analyses}}
  (2025),
\newblock \eprint{2507.11528}.

\bibitem{janssen2008regions}
A.~Janssen and H.~{\"U}nl{\"u},
\newblock \emph{Regions of alternatives with high and low power for
  goodness-of-fit tests},
\newblock Journal of statistical planning and inference \textbf{138}(8), 2526
  (2008).

\end{thebibliography}

\end{document}